\documentclass[12pt]{article}
\usepackage{longtable}
\usepackage{graphicx,amsmath,amsfonts,amssymb,amsthm}

\theoremstyle{plain}

\newtheorem{lemma}{Lemma}

\theoremstyle{remark}
\newtheorem{remark}{Remark}

\begin{document}

\title{\textbf{Transition from the Bose gas to the  Fermi gas through a Nuclear Halo}}

\author{\textbf{V.P. Maslov}
\thanks{ National Research University Higher School of Economics, Moscow, 123458, Russia;
Moscow State  University,   Physics Faculty,  Moscow, 119234, Russia
 }
}
\date{ }

\maketitle

\begin{abstract}
 The  first part  of the paper deals with the behavior of the Bose--Einstein distribution
as the   activity  $a\to 0$.
In particular, the neighborhood of the point
$a=0$
is studied in great detail,
and the expansion of both the Bose distribution
and the Fermi distribution
in powers of the parameter~$a$
is used.

This  approach allows  to  find the value of the parameter~$a_0$,
for which the Bose distribution (in the statistical sense)
becomes zero.

In the  second part of the paper, the process of separation of a nucleon from the atom's nucleus is studied from the mathematical point of view. At the moment when the nucleon tears away from the fermionic nucleus, the nucleus becomes a boson. We investigate the further transformations of
bosonic and fermionic separation states in a small neighborhood of the pressure $P$ equal to zero.
 We use  infinitely small quantities to modify the parastatistical distribution. Our conception  is based on interpolation formulas yielding expansions in powers of the density. This method differs from those in other models based on the interaction potential  between two or three particles.

We obtain new important relations connecting the temperature with the chemical potential during the separation of a nucleon from the atom's nucleus. The obtained relations allow us to construct, on an antipode of sorts of the Hougen--Watson P-Z  chart, the very high temperature isotherms corresponding to nuclear matter.  We call  a new diagram a-Z chart. It is proved mathematically  that the passage of particles satisfying the Fermi--Dirac distribution to the Bose--Einstein distribution in the neighborhood of pressure $P$  equal to zero occurs in a  region known as the ``halo''. 

We have obtained  a new  table  for nuclear  physics  which   demonstrates a monotonic relation between   the nucleus mass number~$A$, the binding energy  $E_b$, the minimum value of activity $a=a_0$,  the chemical  potential $\mu_0=T\log{a_0}$,  the compressibility  factor $F={PV}/{NT}$ for $a_0$,  and  also    the minimum value of  mean square fluctuation  $ \delta \mu_{min}=T/{\delta N}\leq \delta \mu$.  The values $\delta \mu_{min}$  and  $\delta N$  are involved  in   uncertainty relations of nuclear physics.

\end{abstract}

\section{Statistical transition of the Bose gas to the Fermi gas}

\noindent\textbf{1.}
It is well known that a Bose particle
consists of two bound Fermi particles.
 For a Bose particle to become a Fermi particle, it is necessary that
one of the Fermi particles forming the Bose particle
must be ``pushed out'' \emph{beyond} the volume~$V$
under consideration
(in the two-dimensional case,
$V$
is the area)
under the action of some energy.
 In the case of the volume of a ball
(the area of a disk), the radius~$r$
is the radius of the ``shell,'' outside which
the ``entanglement'' of two fermions into one boson
terminates according to the Einstein--Podolsky--Rosen concept.

 Let there be given a macroscopic volume~$V$
of radius at least 1\, mm
filled completely with a Bose gas
(for example, helium-4)
for the activity
$a=1$,
i.e.,
on the caustic.
 We shall determine what amount of energy is required to ``push out''
one fermion from the specific volume under consideration.
 To obtain the total energy,
i.e., the energy needed for the Bose gas
(for example, helium-4) to go over completely
into the Fermi gas
(for example,
helium-3),
we must multiply the resulting quantity
by the number of particles in the gas.
 It is the amount of energy required for the transition of the Bose gas
to the Fermi gas that is the subject of the present study.
 In particular, we shall show that, for the number of degrees of freedom
$D>2$,
we will have a jump of the energy~$E$
at the point
$E=1/\log a$
for
$a=0$,
where
$a$
is the activity,

\begin{remark}
``The barrier'' formed by the shell
is described by a Bardeen--Cooper--Schrieffer type equation
for Cooper pairs~\cite{Masl_Koval_MTN}--\cite{Masl_Koval_TMPH}.
 Since the volume under consideration is macroscopic
(at least 1\, mm$^3$),
by statistical calculations,
the shell far exceeds the volume of the nucleus.
\end{remark}

 In our conception, we use the polylogarithm.
 The polylogarithm is determined by the activity~$a$
and the number of degrees of freedom~$D$,
where
$D=2\gamma+2$
and
$\gamma$
is an auxiliary parameter.
 In the two-dimensional case,
$\gamma=0$,
while,
in the general number theory,
$\gamma\leq 0$.
 The positive value
$\gamma >0$
corresponds to the gas state in classical thermodynamics and
$\gamma <0$,
to the liquid state.

 Both parameters
$a$
and
$D$
are dimensionless,
while the volume~$V$
is a dimensional parameter.
 The dimensionless small parameter
\begin{equation}
\label{hbar}
\frac{\hbar^2}{mVT}
\end{equation}
for
$\gamma=0$
(in the two-dimensional case,
$V$
is the area)
corresponds to the semiclassical approximation
in the sense that the semiclassical asymptotics is expanded
in terms of this parameter.
 Therefore,
as
a rule,
a parameter
$\lambda$
is also added
so that the multiplicative term
before the polylogarithm becomes dimensionless.
 Statistical quantities
divided by the number of particles
i.e., belonging to one particle,
are called \textit{specific quantities}:
$V/N$
is the \textit{specific volume} and
$E/N$
is the \textit{specific energy}.
 We proceed as follows:
first, we shall find the specific energy
and,
further,
we shall multiply it by the number of particles
on the caustic
$a=1$.

 The quantity
$PV/(NT)$,
where
$P$
is the pressure and
$T$
is the temperature,
is dimensionless.
 It is called the \textit{compressibility factor}.
 The temperature~$T$
in the two-dimensional case
is determined from the relation
$M=T \operatorname{Li}_2(a)$,
where
$a=1$,
as
\begin{equation}
\label{t-1}
 T=\sqrt{\frac{M}{\zeta(2)}}\,,
\end{equation}
where
$\zeta(\cdot)$
is the Riemann zeta-function.

 For the number of degrees of freedom
$D>2$,
the density of the Bose gas at the point
$a=0$
vanishes
as
$1/|\log a|$
and
behaves as
$1/\log a$
for the Fermi gas
(see
below),
just as
the energy~$E$
of the Bose gas.
 The coefficient of
$1/\log a$
for the energy of the Bose gas
allows us to determine the energy needed
for ``all'' the Bose particles to go over into
Fermi particles.

 The hidden parameter appearing in the Einstein--Podolsky--Rosen paradox
could unite quantum and classical mechanics
in the sense defined by the author; it can be obtained
from the general arguments in the first chapter of the book~\cite{Landau_SPh-2013}.
 After Sec.~$4$
``Role of Energy,''
the most important notion of time
is introduced in Sec.~$5$
``Statistical Matrix.''
 Since,
in the present paper,
we calculate the statistical energy~$\Delta E$
required for the transition of the Bose gas
into the Fermi gas, it follows that
by using Sec.~$5$
of the book~\cite{Landau_SPh-2013}
we can also calculate the statistical time
\begin{equation}
\label{t-2}
\Delta t \sim \frac{\hbar}{\Delta E}\mspace{2mu},
\end{equation}
given by the authors of~\cite{Landau_SPh-2013}
to explain the transition to macroscopic physics.
 From the point of view of the metamathematical hidden parameter,
this time plays a role similar to that of the mean free time
in classical mechanics.

 Let
$k$
denote the maximal admissible number of particles
that can occupy one energy level,
and let~$N_i$
denote the number of particles
located
at the
$i$th energy level
\footnote{For the transition from a discrete spectrum to a continuous one with respect to the parameter~$V$,
see~\cite{Teor_Vozm},~\cite{Shtern}.}.
 The number~$k$
is the maximal number of particles
located at one energy level of a quantum operator
of the Hamiltonian~$\hat{H}$.
 For a Bose systems,
it is obvious that
$N_i\leq N$.
 Therefore,
for
a Bose system,
$k\leq N$.

 On the other hand, the number
$N_i$
is arbitrarily large, more precisely, maximally large,
in view of the inequality
$N_i\leq N$.
 This implies that
$k=N$,
and
we obtain an equation for~$N$
in which,
on the left-hand
and
right-hand sides,
we have
$N$
from the formula
for the Gentile parastatistics
(see Sec.~4 below).
\medskip

\noindent\textbf{2.}
 To find the energy that is of interest to us
(the energy of transition of Bose particles
to Fermi particles),
it suffices to study the transition
at the point
$a=0$.
 If we wish to widen the interval of the jump from
the Bose distribution to the Fermi distribution,
then we must use parastatistics, or Gentile statistics~\cite{Gentile}
in which the first term in parentheses gives the distribution
for Bose particles and the other term, the parastatistical correction.
 The corresponding formula in the three-dimensional case
is of the form (see~\cite{Kvasn_2})
\begin{equation}
\label{kv-1}
 N =\lambda\bigg\{\int_0^\infty \frac{p^2\, dp}{\exp \{{p^2}/{2m T}\}-1} -
(k+1) \int_0^\infty \frac{p^2\, dp}{\exp\{(k+1){p^2}/{2m T}\} -1}\bigg\}.
\end{equation}

 For the Bose--Einstein distribution
in the case of
$D$
degrees of freedom
the following formulas are valid:
\begin{equation}
\label{BE}
 E=T\Phi ({\gamma+1})Li_{2+\gamma}(a),\qquad N=\Phi Li_{1+\gamma}(a).
\end{equation}
 Here and elsewhere,
$\gamma=D/2-1$,
$T$
is the temperature,
$a=e^{\mu/T}$
is the activity,
$\mu$
is the chemical potential,
$$
\text{
$\Phi=\biggl(\frac{\sqrt{2 \pi m T}}{2 \pi \hbar}\,\biggr)^{2(\gamma+1)}V$,
}
$$
where
$m$
is the mass of one particle,
$V$
is the volume of the system of particles,
and
$\hbar$
is the Planck constant.

 For the Fermi--Dirac distribution,
the following formulas hold:
\begin{equation}
\label{FD}
 E=-T\Phi ({\gamma+1})Li_{2+\gamma}(-a),\qquad N=-\Phi Li_{1+\gamma}(-a).
\end{equation}
 As we see,
formulas~\eqref{BE}
and~\eqref{FD}
differ by sign;
therefore, the activity~$a$
passes through the point
$a=0$.

 In the case of parastatistics,
i.e.,
there are at most
$k$
particles at each level, the following relations hold:
\begin{align}
\label{ep}
 E&={\Phi}T({\gamma+1})
(\text{Li}_{2+\gamma}(a)-\frac{1}{(k+1)^{\gamma+1}}\text{Li}_{2+\gamma}(a^{k+1})),
\\
\label{np}
 N&={\Phi} (Li_{1+\gamma}(a)-\frac{1}{(k+1)^{\gamma}}Li_{1+\gamma}(a^{k+1})).
\end{align}

 According to the conventional definitions,
for
$k=1$,
we have the Fermi case and,
for
$k=\infty$,
the Bose case.

 Let us consider another important dimensionless quantity, which was mentioned above,
namely, the compressibility factor defined by the formula
$Z={PV}/{NT}$.

 The following thermodynamic relation is well known:
\begin{equation}
\label{ez}
 E=-(\gamma+1)\Omega=PV(\gamma+1)
\end{equation}
 As is seen from
relation~\eqref{ez},
$PV$
is expressed in terms of~$E$;
therefore, the compressibility factor can be expressed
in terms of the polylogarithm.
 For the Bose case, the compressibility factor
takes the form
\begin{equation}
\label{ZB}
 Z|_{\text{Bose}}=\frac{Li_{2+\gamma}(a)}{Li_{1+\gamma}(a)}
\end{equation}
and,
for the Fermi case, we have
\begin{equation}
\label{ZF}
 Z|_{\text{Fermi}}=\frac{Li_{2+\gamma}(-a)}{Li_{1+\gamma}(-a)}\mspace{2mu}.
\end{equation}

 For the Gentile statistics
with parameter~$k$,
we have the following expression
for the compressibility factor:
\begin{equation}
\label{ZK}
 Z|_{k}=
\frac{Li_{2+\gamma}(a)-\frac{1}{(k+1)^{\gamma+1}}Li_{2+\gamma}(a^{k+1})}{Li_{1+\gamma}(a)-
\frac{1}{(k+1)^{\gamma}}Li_{1+\gamma}(a^{k+1})}\mspace{2mu}.
\end{equation}
\medskip

\noindent\textbf{3.}
 The compressibility factor~$Z$
multiplied by the temperature~$T$
is the specific energy.
 To obtain the jump of the specific energy,
it suffices to consider the jump of the compressibility factor from
the Fermi system to the Bose system.

 The following representation for the polylogarithm is known:
\begin{equation}
\label{Z}
 Li_{s}(a)=\sum_{i=1}^{\infty}{\frac{a^i}{i^s}}\mspace{2mu};
\end{equation}
substituting this expression into~\eqref{ZB},~\eqref{ZF},~\eqref{ZK},
we obtain the following expansions of the compressibility factor:
\begin{align}
 Z|_{\text{Fermi}}
&=1+a 2^{-\gamma -2}-a^2 2^{-2 \gamma -3} 3^{-\gamma -2} (2^{2 \gamma +4}-3^{\gamma
+2})
\nonumber
\\ &\qquad
-a^3 2^{-3 \gamma -4} 3^{-\gamma -2} (7\cdot 2^{2 \gamma +2}-3^{\gamma +2}-2^{\gamma }
3^{\gamma +3})
\nonumber
\\ &\qquad
-a^4 2^{-4 \gamma -5} 3^{-2 \gamma -3} 5^{-\gamma -2} (2^{4 \gamma +7} 3^{2 \gamma
+3}-2^{4 \gamma +6} 5^{\gamma +2}-3^{2 \gamma +3} 5^{\gamma +2}+2^{2 \gamma +3} 3^{\gamma
+1} 5^{\gamma +3}-2^{\gamma } 3^{2 \gamma +3} 5^{\gamma +3})
\nonumber
\\ &\qquad
+ O(a^5),
\label{z-2}
\\
 Z|_{\text{Bose}}&=1-a 2^{-\gamma -2}-a^2 2^{-2 \gamma -3} 3^{-\gamma -2} (2^{2 \gamma
+4}-3^{\gamma +2})
\nonumber
\\ &\qquad
+a^3 2^{-3 \gamma -4} 3^{-\gamma -2} (7\cdot 2^{2 \gamma +2}-3^{\gamma +2}-2^{\gamma }
3^{\gamma +3})
\nonumber
\\ &\qquad
-a^4 2^{-4 \gamma -5} 3^{-2 \gamma -3} 5^{-\gamma -2} (2^{4 \gamma +7} 3^{2 \gamma
+3}-2^{4 \gamma +6} 5^{\gamma +2}-3^{2 \gamma +3} 5^{\gamma +2}+2^{2 \gamma +3} 3^{\gamma
+1} 5^{\gamma +3}-2^{\gamma } 3^{2 \gamma +3} 5^{\gamma +3})
\nonumber
\\ &\qquad
+ O(a^5).
\label{z-1}
\end{align}

\begin{figure}[h!]
\includegraphics[draft=false]{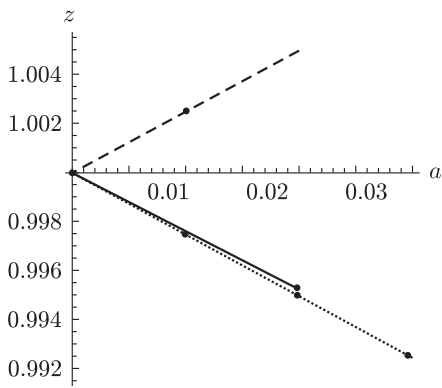}
\caption{Dependence of the compressibility factor~$Z$
on the activity~$a$
for the two-dimensional case.
 Here
$\Phi=100$,
$\gamma=0$.
 The upper
(dashed) curve
corresponds to
the Fermi system.
 The middle
(solid) curve
corresponds to
the Bose system
up to
$O(a^2)$.
 The lower (dotted) curve
corresponds to the exact Bose system.
 The points on the curves
correspond integer~$N$.}
\label{masl_fig_1}
\end{figure}

 The jump of the compressibility factor
is expressed
as
(see Fig.~\ref{masl_fig_1}):
\begin{equation}
\label{dz1}\
\Delta Z(a)= Z|_{K=1}-
 Z|_{\text{Bose}}=\frac{a}{2^{\gamma+1}}+O(a^2),
\end{equation}
while the jump of the \emph{specific} energy
$E_{\text{spec}}$
is of the form
\begin{equation}
\label{de}\
\Delta E_{\text{spec}}(a)=T (\gamma+1)\Delta
 Z(a)=T(\gamma+1)\frac{a}{2^{\gamma+1}}+O(a^2).
\end{equation}

 The dependence of the compressibility factor~$Z$
on the activity~$a$
for the case
$\gamma=0$
is shown in Fig.~\ref{masl_fig_1}.
\medskip

\begin{figure}[h!]
\includegraphics[draft=false]{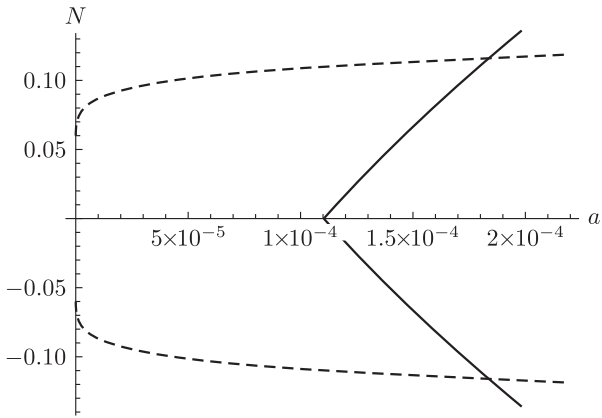}
\caption{Dependence
$N(a)$
for
$W=1000$,
where
$W= V (\lambda^2 T)^{\gamma+1}$,
$\lambda$
is the parameter
depending on the mass, and
$\gamma=0$.
 The dotted curve corresponds to
$N=-1/\log(a)$.
 The sold line
corresponds to
$N=0$.}
\label{masl_fig_2}
\end{figure}

\noindent\textbf{4.}
 As indicated above, the author obtained a self-consistent equation
in which
$N$
is an unknown quantity.
 We are interested in the point at which the number of Bose particles~$N$
is~0,
i.e., the point at which Bose particles vanish.
 An example of this point is given in Fig.~\ref{masl_fig_2}.
 Let us note that the value of the activity~$a$
is not zero at this point;
it depends essentially
on the function~$\Phi$
and the parameter~$\gamma$.
 Let us consider successively the cases of small~$N$,
$\gamma\geq0$,
and
$\gamma<0$.

\textbf{4.1.}
 As pointed out above,
there can be at most
$N$
particles
at each energy level.
 Therefore,
for the Bose system,
we put
$k=N$.
 Equation~\eqref{np} takes the form:
\begin{equation}
\label{np1}
 N= {\Phi} (Li_{1+\gamma}(a)-\frac{1}{(N+1)^{\gamma}}Li_{1+\gamma}(a^{N+1})).
\end{equation}

 Expanding the right-hand side of Eq.~\eqref{np1}
in the small parameter
$N\ll a$,
we obtain
\begin{equation}
\label{N1}
\begin{split}
 N&=\Phi N (\gamma \text{Li}_{\gamma +1}(a)-\log (a) \text{Li}_{\gamma }(a))
\\&\qquad
+\frac{1}{2}W N^2 (\log ^2(a) (-\text{Li}_{\gamma -1}(a))-\gamma ((\gamma +1)
\text{Li}_{\gamma +1}(a)-2 \log (a) \text{Li}_{\gamma
}(a)))+o((\log{a}N)^2).
\end{split}
\end{equation}

\begin{remark}
 We deal with two double limits.
 The first case is
$$
\lim_{a\to 0}a\lim_{N\to 0} N.
$$
 In this case,
$N\ll a$,
and we obtain some value of
$a_{0}$
for which the number of Bose particles of the distribution
$N =\nobreak 0$.

 In the second case,
$N$
is fixed
and
$a\ll N$.
 As a result, the Gentile parastatistical term
vanishes.
\end{remark}

 If,
in the expansion~\eqref{np1},
we take into account the terms up to the second order of smallness
inclusive,
then
we can obtain the equation for~$N$
whose solution is as follows:
\begin{equation}
\label{N2}
 N=2\frac{\gamma \Phi \text{Li}_{\gamma+1}(a)- \Phi \log (a) \text{Li}_\gamma(a)-1}{\Phi
(\log ^2(a) \text{Li}_{\gamma-1}(a)+\gamma ((\gamma+1) \text{Li}_{\gamma+1}(a)-2 \log
(a) \text{Li}_\gamma(a)))}\mspace{2mu}.
\end{equation}

 Equation~\eqref{N2} yields
the following equation for~$a_0$
in the case
$N=0$:
\begin{equation}
\label{a0}
\gamma \Phi \text{Li}_{\gamma+1}(a_0)- \Phi \log (a_0) \text{Li}_\gamma(a_0)-1=0,
\end{equation}

 For small~$a_0$,
the following asymptotic formula holds:
\begin{equation}
\label{a01}
\Phi=-\frac{1}{a_0 \ln{a_0}}\mspace{2mu},
\end{equation}
or
\begin{equation}
\label{a01a}
\Phi=\frac{1}{a_0 \ln{\Phi}}\mspace{2mu}.
\end{equation}

 We obtain
\begin{equation}
\label{dz2}\
\Delta Z(a_0)=\frac{a_0}{2^{\gamma+1}}=-\frac{1}{2^{\gamma+1}\Phi \ln{a_0}}\mspace{2mu}.
\end{equation}

 The maximal number of particles
$N_c$
in the system
is at the caustic point
$a=1$.
 The value of~$N_c$
is given by
\begin{equation}
\label{Nc}
 N_c= {\Phi}(Li_{1+\gamma}(1)-\frac{1}{(N_c+1)^{\gamma}}Li_{1+\gamma}(1)).
\end{equation}
 We shall use the integral representation of the polylogarithm
$\operatorname{Li}$,
obtaining
\begin{equation}
\label{8}
 N_c = \Phi\beta \int_0^\infty \bigg(\frac{1}{\exp\{{\varepsilon}/{T}\}-1} -
\frac{N_c+1}{\exp\{(N_c+1){\varepsilon}/{T}\}-1}\bigg)\, d\varepsilon, \qquad
\beta=1/T.
\end{equation}
\medskip

\textbf{4.2.}
 Consider the case
$D=2$.
 Let
$N=N_c$
be the solution of Eq.~\eqref{8}.
 We shall evaluate the integral
in~\eqref{8}
(with the same integra)
taken from
$\delta$
to~$\infty$
and
then pass to the limit as
$\delta\to0$.
 After making the change
$\beta x=\xi$
in the first term
and
$\beta(N_c+1)x=\xi$
in the second term,
where
$\beta=1/T$,
we
obtain
\begin{align}
 N_c&=\Phi\int_{\delta \beta}^\infty\frac{\,d\xi}{e^\xi-1}
-\Phi\int^\infty_{\delta \beta (N_c+1)}\frac{\,d\xi}{e^\xi-1} +O(\delta)=\Phi
\int^{\delta \beta (N_c+1)}_{\delta \beta}\frac{\,d\xi}{e^\xi-1}+O(\delta)
\label{12}\\&\sim\Phi\int^{\delta \beta (N_c+1)}_{\delta
\beta}\frac{\,d\xi}\xi+O(\delta) =\Phi\{\ln(\delta
\beta (N_c+1))-\ln(\delta \beta)\}+O(\delta)=\Phi\ln (N_c+1) +O(\delta). \label{13}
\end{align}
 In the three-dimensional case,
this formula is of the form~\eqref{kv-1}.

 After passing to the limit as
$\delta\to0$,
we obtain
\begin{equation}
\label{11}
 N_c=\Phi \log (N_c +1).
\end{equation}

 Multiplying the difference of the specific energies
$\Delta E_{\text{spec}}(a_0)$
by the number of particles
defined by formula~\eqref{Nc},
we obtain the difference of the energies
for the whole system of particles
in the units of~$T$
in the case
$\gamma>0$:
\begin{equation}
\label{de1}\
\Delta E=\Delta E_{\text{spec}}(a_0)N_c=\frac{\gamma+1}{2^{\gamma+1}
\ln{\Phi}}Li_{1+\gamma}(1)\biggl(1-\frac{1}{(N_c+1)^{\gamma}}\biggr)
\end{equation}
and,
in the case
$\gamma=0$,
we have
\begin{equation}
\label{de2}\
\Delta E=\Delta E_{\text{spec}}(a_0)N_c=\frac{1}{2 \ln{\Phi}} \log (N_c +1),
\end{equation}
where
$N_c$
can be calculated from formula~\eqref{Nc}
(which contains~$\Phi$).
\medskip

\noindent\textbf{4.3.}
 Let us pass to the case
$\gamma<\gamma_0<0$,
i.e.,
$D<D_0<2$.

 For the case in which
$\gamma<\gamma_0<0$,
the following lemma holds.

\begin{lemma}
 Consider the integral
\begin{equation}
 N=B\int_0^\infty\biggl(\frac{1}{e^{\beta x-\beta \mu}-1}-\frac{k}{e^{k(\beta x-\beta
\mu)}-1}\biggr)x^\gamma \,dx,
\label{1-lem}
\end{equation}
where
$-1< \gamma <\gamma_0<\nobreak 0$, and
$B>0$
and
$k>0$
are constants.

 Then
\begin{equation}
 N=-\frac{B}{\beta^{\gamma+1}}\mspace{2mu}c_{\beta\mu,\gamma}+\frac{B
k^{-\gamma}}{\beta^{\gamma+1}}\mspace{2mu}c_{k \beta\mu,\gamma},
\label{form}
\end{equation}
where
\begin{equation}
c_{\mu,\gamma}=\int_0^\infty\biggl(\frac{1}{\xi-\mu}-\frac{1}{e^{\xi-\mu}-1}\biggr)
\xi^\gamma \,d\xi.
\label{ck}
\end{equation}
\end{lemma}

 By the Lemma, Eq.~\eqref{N1}
takes the form
\begin{equation}
 N_c={\Phi}C(\gamma)(-1+(N_c+1)^{-\gamma}),
\label{form1}
\end{equation}
where
\begin{equation}
 C(\gamma)=\frac{1}{\Gamma(\gamma+1)}\int_0^\infty
\biggl(\frac{1}{\xi}-\frac{1}{e^{\xi}-1}\biggr)\xi^\gamma \,d\xi.
\label{ck-1}
\end{equation}

 As
$N_c\to\infty$,
we can neglect the summand~$-1$
in formula~\eqref{form1} as well as~$1$
compared to~$N_c$,
whence we obtain
\begin{align}
 N_c&={\Phi}C(\gamma)N_c^{-\gamma},
\label{form2}
\\\
N_c&=({\Phi}C(\gamma))^{1/(\gamma+1)}
=T\biggl(\biggl(\frac{\sqrt{2 \pi m}}{2 \pi
\hbar}\biggr)^{2(\gamma+1)}V C(\gamma)\biggr)^{1/(\gamma+1)};
\label{form3}
\end{align}
as a final result, we arrive at the expression
\begin{equation}
 N_c=T(V C(\gamma))^{1/(\gamma+1)} \biggl(\frac{\sqrt{2 \pi m}}{2 \pi
\hbar}\biggr)^{2}.
\label{form4}
\end{equation}
 For the jump of the energy,
we obtain the expression
\begin{equation}
\label{de3}\
\Delta E=\frac{\gamma+1}{2^{\gamma+1}\Phi \ln{\Phi}}
({\Phi}C(\gamma))^{1/(\gamma+1)}=\frac{\gamma+1}{2^{\gamma+1} \ln{\Phi}}
 C(\gamma)^{1/(\gamma+1)}\Phi^{-{\gamma}/({\gamma+1})}.
\end{equation}

\section{Using infinitesimal quantities to describe the separation process of neutrons from atomic  nuclei}

A Bose particle consists 	of two Fermi particles which are linked together by an interaction force or, as it is customary to say at the present time, by an entangling force. 	Sometimes the two fermions constituting the boson are absolutely identical. They cannot be distinguished or numbered. Sometimes the  fermions differ in mass, then they can be distinguished. For example, a proton and an electron are fermions of different mass, and together they constitute a pair -- a boson. Such fermions can be distinguished and numbered, using the mechanism of numeration theory.
While if the masses of the particles approach each other and coincide in the limit, then we cannot distinguish them.			

The passage from the case of distinguishable particles to the case when the particles become undistinguishable leads to other quantum-mechanical formulas and therefore, to  different commutation relations~\cite{Faddeev-Skyrme}.     					
																						
The separation of a fermion from the nucleus and the capture of a fermion is a very rapid process. The question of the possibility of monitoring that transformation as if it occurred in slow motion arises. Mathematically, this means that we must  use such  tools  which may be applied to both fermions and bosons.

 In the present paper we use  infinitely small quantities to modify the parastatistical distribution (the Gentile statistical method). This approach allows to apply the ``antipodå''  of the P-Z chart corresponding to the Van-der-Waals equation of state in thermodynamics, also known as the Hougen--Watson chart. In this antipode, the isotherms under consideration, unlike those in the Van-der-Waals model, correspond to extremely high temperatures, which increase even more when the compressibility factor $Z$  decreases.

 The obtained antipode of the P-Z chart adequately corresponds to the behavior of nuclear matter. We will show in this paper that the passage of particles satisfying the Fermi--Dirac distribution to the Bose--Einstein distribution in the neighborhood of pressure $P$  equal to zero occurs in a  region known as the ``halo''. This region is different for various isotopes and depends on chemical potentials.

The method   developed  in this paper differs from those in other models based on the interaction of two or three particles, such as the Faddeev--Skyrme  model or models  involving the Lennard--Jones potential. This method is based on interpolation formulas yielding expansions in powers of the density. Just as in the Van-der-Waals interpolation formula, the zeroth approximation is given by ideal gas, whereas in models based on potentials the zeroth approximation is given by the zero potential.

 Niels Bohr in 1936  proposed  one of the earliest models of the atomic nucleus
in the framework of the theory of compound nucleus~\cite{Bohr}.
Later,  by Carl Weizs\"acker was  obtained a semi-empirical formula for the binding energy of the atomic nucleus~$E_c$:
\begin{equation}\label{Vaiczekker}
E_c=\alpha A - \beta A^{2/3} - \gamma\frac{\mathcal{Z}^2}{A^{1/3}} - \varepsilon \frac{(A/2-\mathcal{Z})^2}{A}+\delta,
\end{equation}
where
$$
\delta = \left\{
\begin{aligned}
&+\chi A^{-3/4}  & \, \text{for even-even nuclei},\\
\,\,\, &0        & \, \text{for nuclei with odd $A$},\\
&-\chi A^{-3/4}  & \, \text{for odd-odd nuclei},\\
\end{aligned}
\right.
$$
$A$ is the mass number (total number of nucleons) in the nucleus,
$\mathcal{Z}$ is the charge number (number of protons) in the nucleus, and
$\alpha$, $\beta$, $\gamma$, $\varepsilon$, and $\chi$
are parameters obtained by statistical treatment of experimental data.
This formula provides  sufficiently exact values of the binding energies and masses for many nuclei,
which permits using it to analyze different properties of the atomic nucleus.

In this paper  we obtain new important relations between the temperature and the chemical potential
in the process of nucleon separation from the atomic nucleus.
For this, we use the parastatistic relations modified by
the mathematical notion of infinitesimals.

\subsection{New parastatistics}

Let us assume that the energy of each of the particles takes one of the values of the given discrete spectrum. We shall denote by $N_i$ the number of particles located at the $i$-th level energy.

Besides the Bose--Einstein and Fermi--Dirac statistics, physicists use parastatistics (aka Gentile statistics ~\cite{Gentile}) that generalizes the two previously mentioned statistics, which are thus particular cases of parastatistics. According to the latter, as it was said earlier,  at each energy level, there may be no more than $K$ particles. By the usual definitions, the Fermi case is realized for $K=1$. In the case of Bose statistics  $N_i\leq N$ because then
$\sum_{i} N_i = N$, implies that $K=N$.

We will be dealing with infinitely small $K$ and $N$ equal to each other. General approach see in~\cite{Nonstan_anal}.

In the case of parastatistics, we have the following relations, in which the first term in parentheses gives the distribution for Bose particles, and the second term, the parastatistical correction (compare~\eqref{ep}-\eqref{np}):
\begin{equation}\label{ep-2}
E=\frac{V}{\lambda^{D}}T({\gamma+1}) (\operatorname{Li}_{2+\gamma}(a)-\frac{1}{(K+1)^{\gamma+1}}\operatorname{Li}_{2+\gamma}(a^{K+1})),
\end{equation}
\begin{equation}\label{np-2}
N= \frac{V}{\lambda^{D}} (\operatorname{Li}_{1+\gamma}(a)-\frac{1}{(K+1)^{\gamma}}\operatorname{Li}_{1+\gamma}(a^{K+1})),
\end{equation}
where $\operatorname{Li}_{(\cdot)}(\cdot)$  is the polylogarithm function,
$a =e^{\mu/T}$ is the activity ($\mu$ being the chemical potential, $T$ being the temperature),
$\gamma=D/2-1$, $D$  is the number of degrees of freedom (the dimension),
$\lambda$ is the de Broglie wavelength:
\begin{equation}\label{lam}
\lambda=\sqrt{\frac{2\pi\hbar^2}{m T}},
\end{equation}
where $\hbar$ is the Planck constant, $m$ the mass of one particle.

 The de Broglie wavelength is related  to the
Bohr--Sommerfeld quantization condition, i.e., to the quantization of the angular momentum (see~\cite{Litvinov_Mas_dequan}  and related works, in particular~\cite{Tyurin}--\cite{Chen}).

\subsection{Passage from the Bose-type region to the Fermi-type region through a nuclear  halo}

Before completely  separating  from  a nucleus,   a nucleon stays within an area  around a  nucleus where it is bound to  a nucleus  with entanglement  forces.    In this area the probability  to   discover a neutron  is higher  than  the probability to discover  a proton. It is  well-known that the  area where the density  distribution of neutrons  much exceeds     the density  distribution of protons  is called a nuclear halo.

The neutron halo in atomic nuclei is defined as an extended
distribution of neutron density and a narrow momentum distribution
of fragmentation products~\cite{Penion}. Here the uncertainty principle
manifests itself: when the distribution in space is wide, then the
momentum distribution is narrow.

We will call  the area similar to a nuclear halo  a \emph{region of uncertainty}.
As shown in this paper, the width of the  region of uncertainty is determined
by the value of the chemical potential $\mu$: the larger is the
value of the chemical potential, the narrower is the region of uncertainty. For those
elements whose chemical potential is close to zero, the width of the
halo tends to infinity.

This can be explained as follows. The number of particles $N$   is a  conjugate value to the  chemical potential.  For these values
the statistical analogue of  the  Heisenberg uncertainty principle is valid (see~\cite{Gilm}):
\begin{equation}\label{sn}
\Delta \mu_0 \Delta N\ge T.
\end{equation}

 In the  case when  $\Delta\mu_0=-\infty$,  the number of particles  is known.  However  in the  case when  the value $\Delta\mu_0$  is finite,  for number of particles $N$  there exists  a finite uncertainty value   $\Delta N$.  When  $\Delta\mu_0\to0$, the value~$\Delta N$  tends to infinity and corresponds  to  the infinitely wide  region of uncertainty.

The region corresponding to the difference of pressure $P=0$ and to an infinitely small sequence $\{P_K\}\to 0$ constitutes the nuclear halo, or  the region of uncertainty.
 The passage from the Bose-type region to the Fermi-type region occurs through the nuclear halo,  which  contains the value of the pressure $P=0$.
 We denote  by $a_0$  the maximal value of the activity $a$ for Bose particles as $N\to 0$.  The quantity $a_0$ indicates the maximal value of the activity at which the decomposition of bosons into fermion occurs.

 Let us   recall some definitions and relationships  that we used in the paper~\cite{Arxiv_2018}.  It will be helpful  to consider  more general  cases  later  in  this  work.

For an ideal gas of dimension $D=3$, relations  \eqref{ep-2}, \eqref{np-2} become
\begin{equation}\label{Gent}
N=\frac{V}{\lambda^{3}}(\operatorname{Li}_{3/2}(a) -\frac{1}{(K+1)^{1/2}}\operatorname{Li}_{3/2}(a^{K+1})),
\end{equation}
\begin{equation}\label{Mp}
{E}= \frac{3}{2} \frac{{V}}{\lambda^{3}}T(\operatorname{Li}_{5/2}(a)-\frac{1}{(K+1)^{3/2}}\operatorname{Li}_{5/2}(a^{K+1})).
\end{equation}

The expansion of the summand
$\frac{1}{(K+1)^{1/2}}\operatorname{Li}_{3/2}(a^{K+1})$ from formula \eqref{Gent} in small values of $K$ has the form:
\begin{equation}	 \label{raz3}
\begin{split}
&\frac{1}{(K+1)^{1/2}}\operatorname{Li}_{3/2}(a^{K+1})=\operatorname{Li}_{3/2}(a)-[K ( \operatorname{Li}_{3/2}(a)/2-\log (a)\operatorname{Li}_{1/2}(a))+O(K^2).\\
\end{split}
\end{equation}
Let  $B=V/\lambda^3 > 0$. Then equation~\eqref{Gent} for small $K$ acquires the form:
\begin{equation}	 \label{N01-0}
\begin{split}
&N=B K (\frac{1}{2} \operatorname{Li}_{3/2}(a)-\log (a)\operatorname{Li}_{1/2}(a))+O(K^2).\\
\end{split}
\end{equation}

In our considerations   $N$ is an infinitely small number,  ò.å. $N= \alpha_i$,   where a sequence  of  infinitesimals
 $\alpha_i \to 0$ as $i\to  \infty$.
Similarly  $K= \beta_i$,   where   a sequence  of  infinitesimals   $\beta_i\to 0$ as $i \to  \infty$.
The sequences  $\alpha_i$  and   $\beta_i$  are  similar to each other, i.e.
 $\lim_{i\to\infty}{\frac{\alpha_i}{\beta_i}}=1$.
Thus we are not dealing with the Fermi statistics or the Bose statistics, but with a parastatistics of a new type, which can be called  a Bose-like statistics.

Dividing both sides of~\eqref{N01-0} by $N$ and taking the limit as
 $K\to 0$, yields an expression for $a_0$, i.e., the value of $a$ for which $K=N=0$:
\begin{equation}
\label{N=0}
\frac{1}{2} \operatorname{Li}_{3/2}(a_0)-  \log (a_0)\operatorname{Li}_{1/2}(a_0)-B^{-1}=0.
\end{equation}

Equation~\eqref{N=0} in the case of an arbitrary coefficient $\gamma=D/2-1$ instead of $1/2$, after similar arguments, acquires the form:
\begin{equation}
\label{N=0-gen}
\gamma \operatorname{Li}_{\gamma+1}(a_0)-  \log (a_0)\operatorname{Li}_{\gamma}(a_0)-\frac{\lambda^{2(\gamma+1)}}{V}=0.
\end{equation}

Equation~\eqref{N=0-gen} has a unique solution $a_0>0$ that depends on  $B$ and $\gamma$.

In the case $K=N$, equation \eqref{Gent} acquires the form
\begin{equation}\label{GentKN}
	N=B(\operatorname{Li}_{3/2}(a) -\frac{1}{(N+1)^{1/2}}\operatorname{Li}_{3/2}(a^{N+1})).
\end{equation}

This equation obviously has the solution $N\equiv 0$ for any $a\geq 0$. However, for $a>a_0$, it has one more nonnegative solution $N(a)$.
This can be verified by constructing the graphs of the right-hand and left-hand sides of \eqref{GentKN} as a function of $a$ for an arbitrary fixed $N>0$.
The right-hand side of the equation is  zero for $a=0$ and monotonically grows for $a\in(0,\infty)$, while the left hand side is a constant that does not depend on $a$.

Substituting the obtained relation $N(a)$ in formula \eqref{Mp}, we can find the dependence $E(a)$, and with it the pressure $P(a)$, by using the relation
\begin{equation}\label{Epv}
	\vec{E=(\gamma+1)PV.}
\end{equation}

Let us substitute the obtained relation into the graph of the compressibility factor
\begin{equation}\label{Zdef}
	Z=PV/(NT)
\end{equation}
as a function of $P$ (this graph is known as the Hougen--Watson chart).

Now  we place a minus sign before  dependencies  $P(a)$ and $N(a)$   for  the Bose branch, i.e.   these  values are considered as negative.
It follows  from this   that    in the Bose-like region $Z(a)>0$. We will refer to a new chart as the  a-Z chart.

The value   $a=a_0$  is the minimum value of activity for the Bose branch, since for  $a<a_0$ the region of uncertainty   occurs  where $N\equiv0$  and $P\equiv0$  for all  $a$  up to the  value of  $a=0$. Here the passage of particles  to the Fermi  branch  begins.

For Fermi statistics in the case $D=3$, we have the relations
\begin{equation}\label{Nf}
	N=-\frac{{V}}{\lambda^{3}}\operatorname{Li}_{3/2}(-a)  ,
\end{equation}
\begin{equation}\label{Ef}
	E= -\frac{3}{2} \frac{{V}}{\lambda^{3}}T \operatorname{Li}_{5/2}(-a).
\end{equation}

Let us call the curve on the  a-Z chart, constructed according to formulas~\eqref{Nf}--\eqref{Ef} of Fermi statistics,
the Fermi branch. The pressure $P$, as well as the number of particles $N$, on the Fermi branch is positive.

\subsection{a-Z chart  for  nuclear matter}

The value of the activity $a$ for a known value of the temperature $T$ determines the corresponding value of the chemical potential
\begin{equation}
	\label{mu0}
	\mu=T \log(a).
\end{equation}
In particular, for $a=a_0$, the higher the temperature, the smaller is $a_0$, and
the larger is the value of $|\mu_0|$. Thus, as the temperature grows, the point of passage $\mu_0$ approaches the point  $\mu=-\infty$ at which the pressure $P$ changes sign (see~\cite{MTN_98-1}).

Table~\ref{tabl:t1}  in Appendix  presents values  of  $a_0$  and  $\mu_0= T \log a_0$  for isotopes of stable  chemical elements.

The value of $a_0$ in the case of separation can be found by means of formula \eqref{N=0}, taking into account the expression of the de Broglie wavelength
$\lambda$ in terms of the volume $V$ of the nucleus, its temperature $T$ and its mass $m$. The volume of the nucleus is taken to be that of a ball of radius  $r_0=A^{1/3} 1.2\times 10^{-15}$ m$^3$. The temperature $T$ of the nucleus expressed in energy units is taken equal to the binding energy of the nucleus $E_b$, equal to the core temperature $T$ (obtained from the database IsotopeData).

Table~\ref{tabl:t1}    demonstrates a monotonic relation between   the nucleus mass number~$A$, the binding energy  $E_b$, the minimum value of activity $a=a_0$,  the chemical  potential $\mu_0=T\log{a_0}$, and the compressibility  factor $Z={PV}/{NT}$ for $a_0$.

\begin{figure}
	\centering
	\includegraphics[width=8.6cm]{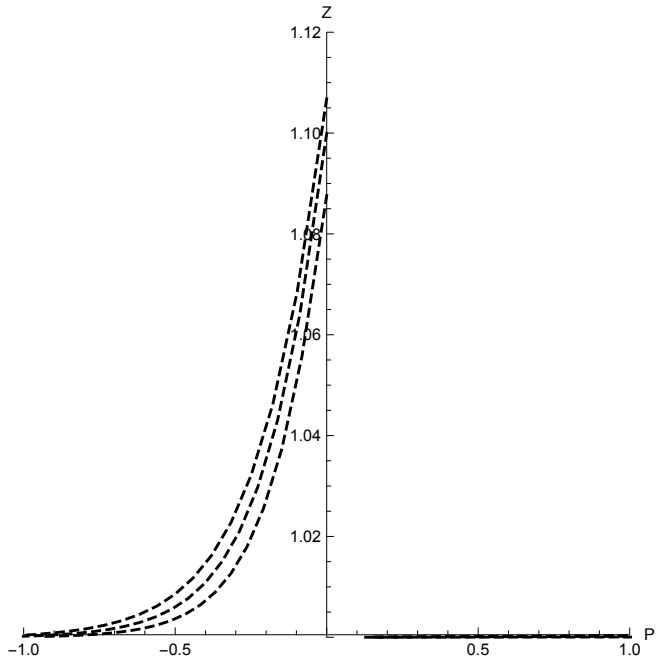}
	\caption{Dependence of the compressibility factor $Z$ on the pressure $P$, expressed in the units MeV/fm $^3$ for carbon-12, nitrogen-14, fluorine-19 (from top to bottom). The continuous line represents the line $Z=1$. It is the beginning of the Fermi branch. The hashed lines show isotherms of  the Bose branch, constructed according to formulas \eqref{Gent}--\eqref{Mp}. The temperature is equal to the nucleus binding energy $E_b$. The corresponding values of $E_b$ and $a_0$ are given in Table~\ref{tabl:t1} in Appendix.}
	\label{fig_01}
\end{figure}

We have obtained an equation~\eqref{N=0} from which we can find the value of $a_0$,
and can determine the temperature $T$ at which this value is attained.

Fig.~\ref{fig_01} shows the dependence of the compressibility factor $Z= PV/(NT)$ on the pressure $P$, expressed in the units MeV/fm$^3$ carbon-12, nitrogen-14, fluorine-19.
The dashed lines are the isotherms of the Bose branch constructed by means of
formulas~\eqref{Gent}, \eqref{Mp}, \eqref{Zdef},  and \eqref{Epv}.  The isotherms    are parametric curves  $P(a)$, $Z(a)$.

\begin{figure}
	\centering
	\includegraphics[width=8.6cm]{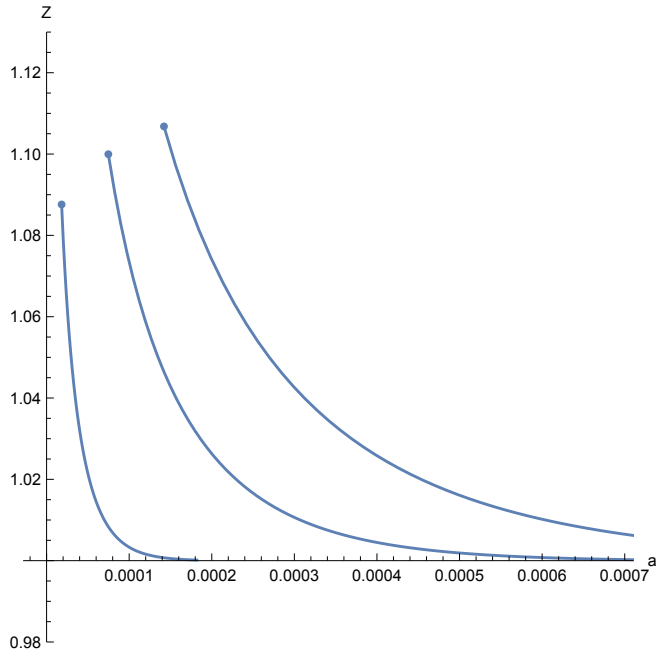}
	\caption{Dependence of the compressibility factor $Z$ on the activity $a$, expressed in the units MeV/fm $^3$ for fluorine-19,  nitrogen-14, carbon-12 (from left to right). The lines show isotherms of  the Bose branch, constructed according to formulas \eqref{Gent}--\eqref{Mp}. The temperature is equal to the nucleus binding energy $E_b$ (see Table~\ref{tabl:t1}  in Appendix).}
	\label{fig_02}
\end{figure}

 \begin{figure}
	\centering
	\includegraphics[width=8.6cm]{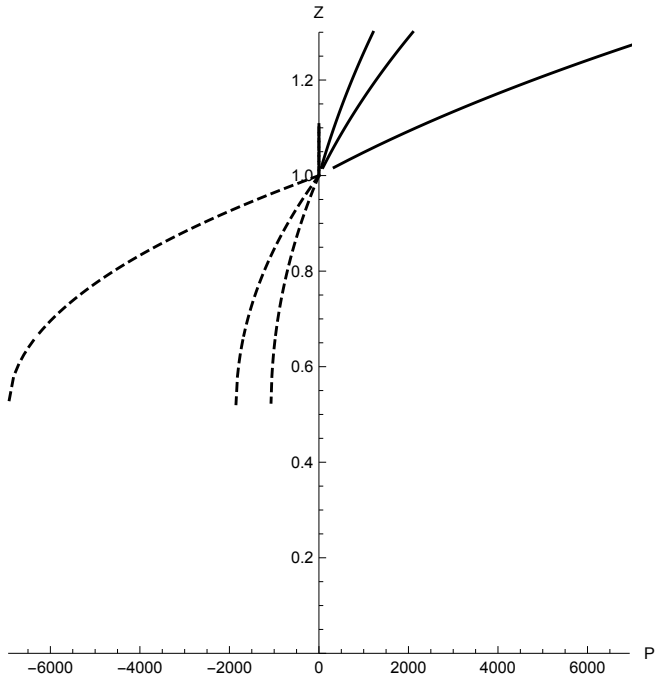}
	\caption{Dependence of the compressibility factor $Z$ on the pressure $P$, expressed in the units MeV/fm $^3$ for fluorine-19,  nitrogen-14, carbon-12   (from left to right in the  region  $P<0$). The continuous line represents the Fermi branch.The hashed lines show isotherms of  the Bose branch, constructed according to formulas \eqref{Gent}--\eqref{Mp}.}
	\label{fig_03}
\end{figure}

Let $\{P_K\}$ be an infinitely small sequence, coinciding with the infinitely small quantity $K$.
 It can be shown that  the corresponding sequence $\{Z_k\}$ tends to  a number  greater than unity   in the Bose-like  region and   to  unity  in the  Fermi-like region (see Fig.~\ref{fig_01}).

The temperature is equal to the extremal value of the binding energy   $E_b$, indicated in Table~\ref{tabl:t1}.
To each value of $a_0$ there corresponds a definite value of the temperature  $T$. In turn, to each value of $T$ there corresponds an isotherm on the Hougen--Watson chart.
These isotherms lie in the second quadrant.
If the volume is constant, the temperature characterizing the isotherm becomes smaller as the point $a=a_0$ becomes nearer to $a=1$.

Thus we can say that the Van-der-Waals isotherms are in a sense opposite to the isotherms of nuclear matter shown in Fig.~\ref{fig_01}.

This shows that the chemical potential $\mu$  at $P=0$ does not become equal to minus infinity and so the  axis  $Z$  at $P=0$ is not the boundary between  two unrelated structures. Since the value of $|\mu_0|$ is very large
but not infinite between the values of the infinitely small quantities
$\{P_K\}$ and the region obeying the Fermi--Dirac distribution, there is a  ``halo'' dividing the Bose region from the Fermi region.

The value of the chemical potential $\mu_0$ determines the halo width.   The halo  expands  as  $\mu_0$  is close to zero. It can become infinitely large.

In the classical thermodynamics this situation  corresponds to passing through the  point where the pressure vanishes (i.e. passing  from  a negative  compressibility factor  $Z$ to a positive $Z$).  The  halo is  a region of  indeterminacy: nobody knows what happens with  the  particles  in the  domain  of a nuclear  halo. We can refer  to halo  as the lacunary indeterminacy.

We do not consider the problem of proving that the isotherms on Fig.~\ref{fig_01} exist. We give an approximate solution of the obtained equations (see similar approaches in  papers~\cite{Bruno},~\cite{Weinstein_Gerbe}), in which all the isotherms corresponding to different values of $a_0$ are approximately constructed.

In the point $a=1$ (i.e. on the spinodal)  the compressibility  factor has  the  form:
   \begin{equation}\label{Zc}
  Z_s=\frac{\zeta(5/2)}{\zeta(3/2)} \frac{1 -\frac{1}{(N_s+1)^{3/2}}}{1 -\frac{1}{(N_s+1)^{1/2}}},
   \end{equation}
where  $N_s$ is determined  from the relation
\begin{equation}\label{np-1}
	N_s= \frac{{V}}{\lambda^{3}}\zeta(3/2) \left(1-\frac{1}{(N_s+1)^{1/2}}\right).
\end{equation}

In the Fig.~\ref{fig_02}  upper ends of isotherms  correspond  to   $a=a_0$.
In the Fig.~\ref{fig_01}  corresponding points  lie  on the axis  $Z$.
This makes it possible    to consider and compare  behavior   of isotherms of different chemical elements.

In Fig.~\ref{fig_03}  isotherms shown in Fig.~\ref{fig_01}  are drawn on a different scale.

One can see in Fig.~\ref{fig_03}  that  the tangent to the lead curve  in  the  bend  point  is  parallel to the axis Z.    This is  a very  important  point where the activity  $a=1$. It connects the  matter under consideration  with the Van-der-Waals  gas. This point is evidence of passing  to the nuclear matter.  Nuclear matter is  a substance  which  in contrast to the nucleus has no boundaries.
 The  bending down of the lead  isotherm  indicates that  in  the separation  process of neutrons  from an atomic  nuclei  we pass  to another substance of nuclear matter.

Thus we have  described  a complete picture  of passing  from a substance which occurs when  a neutron separates from an atomic nucleus to the  nuclear matter substance.

\section*{Conclusion}

We  have shown that  using  infinitesimals $N$  and $K$  equally fast  tending  to zero leads
to the determination of the value of a parameter $a_0$ that allows us to construct an antipode of sorts of the Hougen--Watson P-Z chart for nuclear matter. We call  a new  diagram    the a-Z chart.  We have shown that, knowing the values of $a_0$, we can construct all the isotherms on the a-Z chart.  The smaller  the value of $a_0$ the higher temperature of nuclear matter is.

In the paper it was shown that    during separation  of neutrons from atomic nuclei,  between Bose-like  and  Fermi-like  regions  there exists a  nuclear halo, or a region  of uncertainty related to the  chemical potential   $\mu$. The properties of  halo  may be used to model a medium for filtering radioactive elements. Such a filter turned out to be more efficient that the Darcy filtering medium used in the paper~\cite{Masl_Myas} (also see~\cite{RJMP_25-3}).

The obtained in the paper  rigorous mathematical approach unexpectedly yields a new picture, which adequately describes   the passing to the   nuclear matter during   separation  process of  neutrons from atomic nuclei.  This led to  a new  table for nuclear  physics (see Appendix)   which   demonstrates a monotonic relation between   the nucleus mass number~$A$, the binding energy  $E_b$, the minimum value of activity $a=a_0$,  the chemical  potential $\mu_0=T\log{a_0}$, and the compressibility  factor $Z={PV}/{NT}$ for $a_0$ and  also
the minimum value of  mean square fluctuation  $ \delta \mu_{min}=T/{\delta N}\leq \delta \mu$.  The values $\delta \mu_{min}$  and  $\delta N$  are involved  in   uncertainty relations of nuclear physics.

\section*{Additional information}

\textbf{Competing  interests:}  The author declares no competing  interests.

\newpage

\appendix

\section*{Appendix}

Table~\ref{tabl:t1}  is based  on the   IsotopeData which  contains  256 stable elements.
The table does not include 2 elements:
hydrogen-1 and lithium-3.

\begin{longtable}{|c|c|c|c|c|c|c|c|c|}
	\caption{Parameters for  stable isotopes of various  nuclei} \label{tabl:t1} \\
	\hline
	$\mathcal{N}$&nucleus&$\mathcal{Z}$& $E_b$,\,MeV  &$\delta N$&$\delta\mu_{min} $ & $a_0$& $\mu_0$,\,MeV&$F(a_0)$ \\
	\hline
	\endfirsthead
	\hline
	$\mathcal{N}$&nucleus&$\mathcal{Z}$& $E_b$,\,MeV  &$\delta N$&$\delta\mu_{min}$ & $a_0$& $\mu_0$,\,MeV&$F(a_0)$ \\
	\hline
	\endhead
	\hline
	\endfoot
	\hline
	\endlastfoot
		1   & lead-208         & 82 & 1.636*$10^{3}$ & 2.9672*$10^{-1}$ & 5.515*$10^{3}$  & 6.57*$10^{-10}$ & -3.46*$10^{4}$ & 1.046203 \\
		2   & lead-207         & 82 & 1.629*$10^{3}$ & 2.9685*$10^{-1}$ & 5.488*$10^{3}$  & 6.7*$10^{-10}$  & -3.44*$10^{4}$ & 1.046245 \\
		3   & lead-206         & 82 & 1.622*$10^{3}$ & 2.9697*$10^{-1}$ & 5.463*$10^{3}$  & 6.83*$10^{-10}$ & -3.42*$10^{4}$ & 1.046286 \\
		4   & thallium-205     & 81 & 1.615*$10^{3}$ & 2.971*$10^{-1}$  & 5.436*$10^{3}$  & 6.96*$10^{-10}$ & -3.41*$10^{4}$ & 1.046328 \\
		5   & mercury-204      & 80 & 1.609*$10^{3}$ & 2.9723*$10^{-1}$ & 5.412*$10^{3}$  & 7.1*$10^{-10}$  & -3.39*$10^{4}$ & 1.046369 \\
		6   & lead-204         & 82 & 1.608*$10^{3}$ & 2.9723*$10^{-1}$ & 5.408*$10^{3}$  & 7.11*$10^{-10}$ & -3.39*$10^{4}$ & 1.046372 \\
		7   & thallium-203     & 81 & 1.601*$10^{3}$ & 2.9736*$10^{-1}$ & 5.384*$10^{3}$  & 7.25*$10^{-10}$ & -3.37*$10^{4}$ & 1.046414 \\
		8   & mercury-202      & 80 & 1.595*$10^{3}$ & 2.9748*$10^{-1}$ & 5.362*$10^{3}$  & 7.38*$10^{-10}$ & -3.35*$10^{4}$ & 1.046454 \\
		9   & mercury-201      & 80 & 1.587*$10^{3}$ & 2.9762*$10^{-1}$ & 5.334*$10^{3}$  & 7.54*$10^{-10}$ & -3.33*$10^{4}$ & 1.046498 \\
		10  & mercury-200      & 80 & 1.581*$10^{3}$ & 2.9774*$10^{-1}$ & 5.311*$10^{3}$  & 7.68*$10^{-10}$ & -3.32*$10^{4}$ & 1.04654  \\
		11  & mercury-199      & 80 & 1.573*$10^{3}$ & 2.9788*$10^{-1}$ & 5.281*$10^{3}$  & 7.85*$10^{-10}$ & -3.3*$10^{4}$  & 1.046586 \\
		12  & platinum-198     & 78 & 1.567*$10^{3}$ & 2.9801*$10^{-1}$ & 5.258*$10^{3}$  & 8.*$10^{-10}$   & -3.28*$10^{4}$ & 1.046628 \\
		13  & mercury-198      & 80 & 1.566*$10^{3}$ & 2.9802*$10^{-1}$ & 5.256*$10^{3}$  & 8.*$10^{-10}$   & -3.28*$10^{4}$ & 1.046629 \\
		14  & gold-197         & 79 & 1.559*$10^{3}$ & 2.9815*$10^{-1}$ & 5.23*$10^{3}$   & 8.17*$10^{-10}$ & -3.26*$10^{4}$ & 1.046673 \\
		15  & platinum-196     & 78 & 1.554*$10^{3}$ & 2.9828*$10^{-1}$ & 5.209*$10^{3}$  & 8.33*$10^{-10}$ & -3.25*$10^{4}$ & 1.046715 \\
		16  & mercury-196      & 80 & 1.551*$10^{3}$ & 2.9829*$10^{-1}$ & 5.2*$10^{3}$    & 8.35*$10^{-10}$ & -3.24*$10^{4}$ & 1.046721 \\
		17  & platinum-195     & 78 & 1.546*$10^{3}$ & 2.9842*$10^{-1}$ & 5.18*$10^{3}$   & 8.51*$10^{-10}$ & -3.23*$10^{4}$ & 1.046762 \\
		18  & platinum-194     & 78 & 1.540*$10^{3}$ & 2.9855*$10^{-1}$ & 5.157*$10^{3}$  & 8.68*$10^{-10}$ & -3.21*$10^{4}$ & 1.046805 \\
		19  & iridium-193      & 77 & 1.532*$10^{3}$ & 2.9869*$10^{-1}$ & 5.129*$10^{3}$  & 8.86*$10^{-10}$ & -3.19*$10^{4}$ & 1.046852 \\
		20  & osmium-192       & 76 & 1.526*$10^{3}$ & 2.9882*$10^{-1}$ & 5.107*$10^{3}$  & 9.04*$10^{-10}$ & -3.18*$10^{4}$ & 1.046895 \\
		21  & platinum-192     & 78 & 1.525*$10^{3}$ & 2.9883*$10^{-1}$ & 5.103*$10^{3}$  & 9.05*$10^{-10}$ & -3.18*$10^{4}$ & 1.046898 \\
		22  & iridium-191      & 77 & 1.518*$10^{3}$ & 2.9897*$10^{-1}$ & 5.078*$10^{3}$  & 9.24*$10^{-10}$ & -3.16*$10^{4}$ & 1.046944 \\
		23  & osmium-190       & 76 & 1.513*$10^{3}$ & 2.991*$10^{-1}$  & 5.058*$10^{3}$  & 9.42*$10^{-10}$ & -3.14*$10^{4}$ & 1.046986 \\
		24  & osmium-189       & 76 & 1.505*$10^{3}$ & 2.9924*$10^{-1}$ & 5.029*$10^{3}$  & 9.63*$10^{-10}$ & -3.12*$10^{4}$ & 1.047035 \\
		25  & osmium-188       & 76 & 1.499*$10^{3}$ & 2.9938*$10^{-1}$ & 5.007*$10^{3}$  & 9.83*$10^{-10}$ & -3.11*$10^{4}$ & 1.047079 \\
		26  & osmium-187       & 76 & 1.491*$10^{3}$ & 2.9953*$10^{-1}$ & 4.978*$10^{3}$  & 1.*$10^{-9}$    & -3.09*$10^{4}$ & 1.047129 \\
		27  & tungsten-186     & 74 & 1.486*$10^{3}$ & 2.9966*$10^{-1}$ & 4.959*$10^{3}$  & 1.02*$10^{-9}$  & -3.08*$10^{4}$ & 1.047173 \\
		28  & rhenium-185      & 75 & 1.478*$10^{3}$ & 2.9981*$10^{-1}$ & 4.931*$10^{3}$  & 1.05*$10^{-9}$  & -3.06*$10^{4}$ & 1.047222 \\
		29  & tungsten-184     & 74 & 1.473*$10^{3}$ & 2.9994*$10^{-1}$ & 4.911*$10^{3}$  & 1.07*$10^{-9}$  & -3.04*$10^{4}$ & 1.047267 \\
		30  & osmium-184       & 76 & 1.470*$10^{3}$ & 2.9996*$10^{-1}$ & 4.9*$10^{3}$    & 1.07*$10^{-9}$  & -3.04*$10^{4}$ & 1.047274 \\
		31  & tungsten-183     & 74 & 1.466*$10^{3}$ & 3.0009*$10^{-1}$ & 4.884*$10^{3}$  & 1.09*$10^{-9}$  & -3.02*$10^{4}$ & 1.047316 \\
		32  & tungsten-182     & 74 & 1.459*$10^{3}$ & 3.0023*$10^{-1}$ & 4.861*$10^{3}$  & 1.12*$10^{-9}$  & -3.01*$10^{4}$ & 1.047363 \\
		33  & tantalum-181     & 73 & 1.452*$10^{3}$ & 3.0038*$10^{-1}$ & 4.835*$10^{3}$  & 1.14*$10^{-9}$  & -2.99*$10^{4}$ & 1.047413 \\
		34  & hafnium-180      & 72 & 1.446*$10^{3}$ & 3.0052*$10^{-1}$ & 4.813*$10^{3}$  & 1.17*$10^{-9}$  & -2.98*$10^{4}$ & 1.04746  \\
		35  & tungsten-180     & 74 & 1.445*$10^{3}$ & 3.0054*$10^{-1}$ & 4.807*$10^{3}$  & 1.17*$10^{-9}$  & -2.97*$10^{4}$ & 1.047465 \\
		36  & hafnium-179      & 72 & 1.439*$10^{3}$ & 3.0068*$10^{-1}$ & 4.786*$10^{3}$  & 1.19*$10^{-9}$  & -2.96*$10^{4}$ & 1.047512 \\
		37  & hafnium-178      & 72 & 1.433*$10^{3}$ & 3.0082*$10^{-1}$ & 4.763*$10^{3}$  & 1.22*$10^{-9}$  & -2.94*$10^{4}$ & 1.04756  \\
		38  & hafnium-177      & 72 & 1.425*$10^{3}$ & 3.0098*$10^{-1}$ & 4.735*$10^{3}$  & 1.25*$10^{-9}$  & -2.92*$10^{4}$ & 1.047612 \\
		39  & ytterbium-176    & 70 & 1.419*$10^{3}$ & 3.0112*$10^{-1}$ & 4.713*$10^{3}$  & 1.27*$10^{-9}$  & -2.91*$10^{4}$ & 1.047661 \\
		40  & hafnium-176      & 72 & 1.419*$10^{3}$ & 3.0113*$10^{-1}$ & 4.712*$10^{3}$  & 1.27*$10^{-9}$  & -2.91*$10^{4}$ & 1.047662 \\
		41  & lutetium-175     & 71 & 1.412*$10^{3}$ & 3.0128*$10^{-1}$ & 4.687*$10^{3}$  & 1.3*$10^{-9}$   & -2.89*$10^{4}$ & 1.047713 \\
		42  & ytterbium-174    & 70 & 1.407*$10^{3}$ & 3.0143*$10^{-1}$ & 4.666*$10^{3}$  & 1.33*$10^{-9}$  & -2.87*$10^{4}$ & 1.047762 \\
		43  & ytterbium-173    & 70 & 1.399*$10^{3}$ & 3.0159*$10^{-1}$ & 4.639*$10^{3}$  & 1.36*$10^{-9}$  & -2.86*$10^{4}$ & 1.047815 \\
		44  & ytterbium-172    & 70 & 1.393*$10^{3}$ & 3.0174*$10^{-1}$ & 4.616*$10^{3}$  & 1.39*$10^{-9}$  & -2.84*$10^{4}$ & 1.047867 \\
		45  & ytterbium-171    & 70 & 1.385*$10^{3}$ & 3.0191*$10^{-1}$ & 4.587*$10^{3}$  & 1.43*$10^{-9}$  & -2.82*$10^{4}$ & 1.047923 \\
		46  & erbium-170       & 68 & 1.379*$10^{3}$ & 3.0206*$10^{-1}$ & 4.566*$10^{3}$  & 1.46*$10^{-9}$  & -2.81*$10^{4}$ & 1.047973 \\
		47  & ytterbium-170    & 70 & 1.378*$10^{3}$ & 3.0206*$10^{-1}$ & 4.562*$10^{3}$  & 1.46*$10^{-9}$  & -2.8*$10^{4}$  & 1.047975 \\
		48  & thulium-169      & 69 & 1.371*$10^{3}$ & 3.0222*$10^{-1}$ & 4.538*$10^{3}$  & 1.5*$10^{-9}$   & -2.79*$10^{4}$ & 1.048029 \\
		49  & erbium-168       & 68 & 1.366*$10^{3}$ & 3.0237*$10^{-1}$ & 4.517*$10^{3}$  & 1.53*$10^{-9}$  & -2.77*$10^{4}$ & 1.04808  \\
		50  & ytterbium-168    & 70 & 1.363*$10^{3}$ & 3.024*$10^{-1}$  & 4.507*$10^{3}$  & 1.53*$10^{-9}$  & -2.77*$10^{4}$ & 1.048088 \\
		51  & erbium-167       & 68 & 1.358*$10^{3}$ & 3.0254*$10^{-1}$ & 4.489*$10^{3}$  & 1.57*$10^{-9}$  & -2.75*$10^{4}$ & 1.048137 \\
		52  & erbium-166       & 68 & 1.352*$10^{3}$ & 3.0271*$10^{-1}$ & 4.465*$10^{3}$  & 1.6*$10^{-9}$   & -2.74*$10^{4}$ & 1.048191 \\
		53  & holmium-165      & 67 & 1.344*$10^{3}$ & 3.0287*$10^{-1}$ & 4.438*$10^{3}$  & 1.64*$10^{-9}$  & -2.72*$10^{4}$ & 1.048248 \\
		54  & dysprosium-164   & 66 & 1.338*$10^{3}$ & 3.0304*$10^{-1}$ & 4.415*$10^{3}$  & 1.68*$10^{-9}$  & -2.7*$10^{4}$  & 1.048302 \\
		55  & erbium-164       & 68 & 1.336*$10^{3}$ & 3.0305*$10^{-1}$ & 4.41*$10^{3}$   & 1.69*$10^{-9}$  & -2.7*$10^{4}$  & 1.048306 \\
		56  & dysprosium-163   & 66 & 1.330*$10^{3}$ & 3.0321*$10^{-1}$ & 4.388*$10^{3}$  & 1.72*$10^{-9}$  & -2.68*$10^{4}$ & 1.04836  \\
		57  & dysprosium-162   & 66 & 1.324*$10^{3}$ & 3.0337*$10^{-1}$ & 4.365*$10^{3}$  & 1.77*$10^{-9}$  & -2.67*$10^{4}$ & 1.048416 \\
		58  & erbium-162       & 68 & 1.321*$10^{3}$ & 3.034*$10^{-1}$  & 4.353*$10^{3}$  & 1.77*$10^{-9}$  & -2.66*$10^{4}$ & 1.048425 \\
		59  & dysprosium-161   & 66 & 1.316*$10^{3}$ & 3.0355*$10^{-1}$ & 4.335*$10^{3}$  & 1.81*$10^{-9}$  & -2.65*$10^{4}$ & 1.048477 \\
		60  & dysprosium-160   & 66 & 1.309*$10^{3}$ & 3.0372*$10^{-1}$ & 4.311*$10^{3}$  & 1.86*$10^{-9}$  & -2.63*$10^{4}$ & 1.048534 \\
		61  & gadolinium-160   & 64 & 1.309*$10^{3}$ & 3.0372*$10^{-1}$ & 4.311*$10^{3}$  & 1.86*$10^{-9}$  & -2.63*$10^{4}$ & 1.048534 \\
		62  & terbium-159      & 65 & 1.302*$10^{3}$ & 3.039*$10^{-1}$  & 4.284*$10^{3}$  & 1.9*$10^{-9}$   & -2.61*$10^{4}$ & 1.048594 \\
		63  & gadolinium-158   & 64 & 1.296*$10^{3}$ & 3.0407*$10^{-1}$ & 4.262*$10^{3}$  & 1.95*$10^{-9}$  & -2.6*$10^{4}$  & 1.04865  \\
		64  & dysprosium-158   & 66 & 1.294*$10^{3}$ & 3.0408*$10^{-1}$ & 4.256*$10^{3}$  & 1.96*$10^{-9}$  & -2.59*$10^{4}$ & 1.048656 \\
		65  & gadolinium-157   & 64 & 1.288*$10^{3}$ & 3.0425*$10^{-1}$ & 4.233*$10^{3}$  & 2.*$10^{-9}$    & -2.58*$10^{4}$ & 1.048713 \\
		66  & gadolinium-156   & 64 & 1.282*$10^{3}$ & 3.0443*$10^{-1}$ & 4.21*$10^{3}$   & 2.05*$10^{-9}$  & -2.56*$10^{4}$ & 1.048771 \\
		67  & dysprosium-156   & 66 & 1.278*$10^{3}$ & 3.0446*$10^{-1}$ & 4.198*$10^{3}$  & 2.06*$10^{-9}$  & -2.56*$10^{4}$ & 1.048782 \\
		68  & gadolinium-155   & 64 & 1.273*$10^{3}$ & 3.0462*$10^{-1}$ & 4.179*$10^{3}$  & 2.11*$10^{-9}$  & -2.54*$10^{4}$ & 1.048837 \\
		69  & samarium-154     & 62 & 1.267*$10^{3}$ & 3.0479*$10^{-1}$ & 4.157*$10^{3}$  & 2.16*$10^{-9}$  & -2.53*$10^{4}$ & 1.048896 \\
		70  & gadolinium-154   & 64 & 1.267*$10^{3}$ & 3.048*$10^{-1}$  & 4.156*$10^{3}$  & 2.16*$10^{-9}$  & -2.53*$10^{4}$ & 1.048896 \\
		71  & europium-153     & 63 & 1.259*$10^{3}$ & 3.0498*$10^{-1}$ & 4.128*$10^{3}$  & 2.22*$10^{-9}$  & -2.51*$10^{4}$ & 1.04896  \\
		72  & samarium-152     & 62 & 1.253*$10^{3}$ & 3.0516*$10^{-1}$ & 4.106*$10^{3}$  & 2.28*$10^{-9}$  & -2.49*$10^{4}$ & 1.049019 \\
		73  & europium-151     & 63 & 1.244*$10^{3}$ & 3.0536*$10^{-1}$ & 4.074*$10^{3}$  & 2.34*$10^{-9}$  & -2.47*$10^{4}$ & 1.049088 \\
		74  & samarium-150     & 62 & 1.239*$10^{3}$ & 3.0553*$10^{-1}$ & 4.056*$10^{3}$  & 2.4*$10^{-9}$   & -2.46*$10^{4}$ & 1.049146 \\
		75  & samarium-149     & 62 & 1.231*$10^{3}$ & 3.0573*$10^{-1}$ & 4.027*$10^{3}$  & 2.47*$10^{-9}$  & -2.44*$10^{4}$ & 1.049213 \\
		76  & neodymium-148    & 60 & 1.225*$10^{3}$ & 3.0591*$10^{-1}$ & 4.005*$10^{3}$  & 2.53*$10^{-9}$  & -2.42*$10^{4}$ & 1.049275 \\
		77  & neodymium-146    & 60 & 1.212*$10^{3}$ & 3.0628*$10^{-1}$ & 3.958*$10^{3}$  & 2.67*$10^{-9}$  & -2.39*$10^{4}$ & 1.049402 \\
		78  & neodymium-145    & 60 & 1.205*$10^{3}$ & 3.0648*$10^{-1}$ & 3.931*$10^{3}$  & 2.74*$10^{-9}$  & -2.38*$10^{4}$ & 1.04947  \\
		79  & samarium-144     & 62 & 1.196*$10^{3}$ & 3.067*$10^{-1}$  & 3.899*$10^{3}$  & 2.83*$10^{-9}$  & -2.35*$10^{4}$ & 1.049544 \\
		80  & neodymium-143    & 60 & 1.191*$10^{3}$ & 3.0687*$10^{-1}$ & 3.882*$10^{3}$  & 2.9*$10^{-9}$   & -2.34*$10^{4}$ & 1.049604 \\
		81  & cerium-142       & 58 & 1.185*$10^{3}$ & 3.0706*$10^{-1}$ & 3.86*$10^{3}$   & 2.97*$10^{-9}$  & -2.33*$10^{4}$ & 1.049669 \\
		82  & neodymium-142    & 60 & 1.185*$10^{3}$ & 3.0706*$10^{-1}$ & 3.86*$10^{3}$   & 2.97*$10^{-9}$  & -2.33*$10^{4}$ & 1.049669 \\
		83  & praseodymium-141 & 59 & 1.178*$10^{3}$ & 3.0727*$10^{-1}$ & 3.834*$10^{3}$  & 3.06*$10^{-9}$  & -2.31*$10^{4}$ & 1.049739 \\
		84  & cerium-140       & 58 & 1.173*$10^{3}$ & 3.0745*$10^{-1}$ & 3.814*$10^{3}$  & 3.14*$10^{-9}$  & -2.3*$10^{4}$  & 1.049803 \\
		85  & lanthanum-139    & 57 & 1.165*$10^{3}$ & 3.0767*$10^{-1}$ & 3.785*$10^{3}$  & 3.23*$10^{-9}$  & -2.28*$10^{4}$ & 1.049877 \\
		86  & barium-138       & 56 & 1.158*$10^{3}$ & 3.0787*$10^{-1}$ & 3.762*$10^{3}$  & 3.33*$10^{-9}$  & -2.26*$10^{4}$ & 1.049946 \\
		87  & cerium-138       & 58 & 1.156*$10^{3}$ & 3.0789*$10^{-1}$ & 3.755*$10^{3}$  & 3.34*$10^{-9}$  & -2.26*$10^{4}$ & 1.049953 \\
		88  & barium-137       & 56 & 1.150*$10^{3}$ & 3.0809*$10^{-1}$ & 3.732*$10^{3}$  & 3.43*$10^{-9}$  & -2.24*$10^{4}$ & 1.050023 \\
		89  & barium-136       & 56 & 1.143*$10^{3}$ & 3.083*$10^{-1}$  & 3.707*$10^{3}$  & 3.53*$10^{-9}$  & -2.22*$10^{4}$ & 1.050095 \\
		90  & xenon-136        & 54 & 1.142*$10^{3}$ & 3.0831*$10^{-1}$ & 3.704*$10^{3}$  & 3.53*$10^{-9}$  & -2.22*$10^{4}$ & 1.050098 \\
		91  & cerium-136       & 58 & 1.139*$10^{3}$ & 3.0834*$10^{-1}$ & 3.693*$10^{3}$  & 3.55*$10^{-9}$  & -2.22*$10^{4}$ & 1.050109 \\
		92  & barium-135       & 56 & 1.134*$10^{3}$ & 3.0854*$10^{-1}$ & 3.674*$10^{3}$  & 3.64*$10^{-9}$  & -2.2*$10^{4}$  & 1.050176 \\
		93  & xenon-134        & 54 & 1.127*$10^{3}$ & 3.0875*$10^{-1}$ & 3.652*$10^{3}$  & 3.75*$10^{-9}$  & -2.19*$10^{4}$ & 1.050247 \\
		94  & barium-134       & 56 & 1.127*$10^{3}$ & 3.0875*$10^{-1}$ & 3.649*$10^{3}$  & 3.75*$10^{-9}$  & -2.19*$10^{4}$ & 1.050249 \\
		95  & cesium-133       & 55 & 1.119*$10^{3}$ & 3.0898*$10^{-1}$ & 3.62*$10^{3}$   & 3.87*$10^{-9}$  & -2.17*$10^{4}$ & 1.050328 \\
		96  & xenon-132        & 54 & 1.112*$10^{3}$ & 3.0919*$10^{-1}$ & 3.598*$10^{3}$  & 3.98*$10^{-9}$  & -2.15*$10^{4}$ & 1.050401 \\
		97  & barium-132       & 56 & 1.110*$10^{3}$ & 3.0922*$10^{-1}$ & 3.59*$10^{3}$   & 4.*$10^{-9}$    & -2.15*$10^{4}$ & 1.050409 \\
		98  & xenon-131        & 54 & 1.104*$10^{3}$ & 3.0943*$10^{-1}$ & 3.566*$10^{3}$  & 4.12*$10^{-9}$  & -2.13*$10^{4}$ & 1.050484 \\
		99  & xenon-130        & 54 & 1.097*$10^{3}$ & 3.0965*$10^{-1}$ & 3.542*$10^{3}$  & 4.24*$10^{-9}$  & -2.11*$10^{4}$ & 1.05056  \\
		100 & barium-130       & 56 & 1.093*$10^{3}$ & 3.097*$10^{-1}$  & 3.528*$10^{3}$  & 4.27*$10^{-9}$  & -2.11*$10^{4}$ & 1.050575 \\
		101 & xenon-129        & 54 & 1.088*$10^{3}$ & 3.099*$10^{-1}$  & 3.51*$10^{3}$   & 4.39*$10^{-9}$  & -2.09*$10^{4}$ & 1.050646 \\
		102 & xenon-128        & 54 & 1.081*$10^{3}$ & 3.1013*$10^{-1}$ & 3.485*$10^{3}$  & 4.52*$10^{-9}$  & -2.08*$10^{4}$ & 1.050725 \\
		103 & iodine-127       & 53 & 1.073*$10^{3}$ & 3.1037*$10^{-1}$ & 3.456*$10^{3}$  & 4.67*$10^{-9}$  & -2.06*$10^{4}$ & 1.050809 \\
		104 & tellurium-126    & 52 & 1.066*$10^{3}$ & 3.1059*$10^{-1}$ & 3.433*$10^{3}$  & 4.81*$10^{-9}$  & -2.04*$10^{4}$ & 1.050886 \\
		105 & xenon-126        & 54 & 1.064*$10^{3}$ & 3.1062*$10^{-1}$ & 3.425*$10^{3}$  & 4.83*$10^{-9}$  & -2.04*$10^{4}$ & 1.050896 \\
		106 & tellurium-125    & 52 & 1.057*$10^{3}$ & 3.1085*$10^{-1}$ & 3.401*$10^{3}$  & 4.98*$10^{-9}$  & -2.02*$10^{4}$ & 1.050976 \\
		107 & tellurium-124    & 52 & 1.051*$10^{3}$ & 3.1108*$10^{-1}$ & 3.377*$10^{3}$  & 5.14*$10^{-9}$  & -2.01*$10^{4}$ & 1.051057 \\
		108 & tin-124          & 50 & 1.050*$10^{3}$ & 3.1109*$10^{-1}$ & 3.375*$10^{3}$  & 5.15*$10^{-9}$  & -2.*$10^{4}$   & 1.051059 \\
		109 & xenon-124        & 54 & 1.046*$10^{3}$ & 3.1113*$10^{-1}$ & 3.363*$10^{3}$  & 5.17*$10^{-9}$  & -2.*$10^{4}$   & 1.051074 \\
		110 & antimony-123     & 51 & 1.042*$10^{3}$ & 3.1134*$10^{-1}$ & 3.347*$10^{3}$  & 5.32*$10^{-9}$  & -1.99*$10^{4}$ & 1.051146 \\
		111 & tin-122          & 50 & 1.036*$10^{3}$ & 3.1158*$10^{-1}$ & 3.323*$10^{3}$  & 5.49*$10^{-9}$  & -1.97*$10^{4}$ & 1.051229 \\
		112 & tellurium-122    & 52 & 1.034*$10^{3}$ & 3.1159*$10^{-1}$ & 3.32*$10^{3}$   & 5.5*$10^{-9}$   & -1.97*$10^{4}$ & 1.051234 \\
		113 & antimony-121     & 51 & 1.026*$10^{3}$ & 3.1185*$10^{-1}$ & 3.291*$10^{3}$  & 5.69*$10^{-9}$  & -1.95*$10^{4}$ & 1.051323 \\
		114 & tin-120          & 50 & 1.021*$10^{3}$ & 3.1208*$10^{-1}$ & 3.27*$10^{3}$   & 5.87*$10^{-9}$  & -1.93*$10^{4}$ & 1.051404 \\
		115 & tellurium-120    & 52 & 1.017*$10^{3}$ & 3.1212*$10^{-1}$ & 3.259*$10^{3}$  & 5.9*$10^{-9}$   & -1.93*$10^{4}$ & 1.051418 \\
		116 & tin-119          & 50 & 1.011*$10^{3}$ & 3.1236*$10^{-1}$ & 3.238*$10^{3}$  & 6.08*$10^{-9}$  & -1.91*$10^{4}$ & 1.0515   \\
		117 & tin-118          & 50 & 1.005*$10^{3}$ & 3.126*$10^{-1}$  & 3.215*$10^{3}$  & 6.29*$10^{-9}$  & -1.9*$10^{4}$  & 1.051586 \\
		118 & tin-117          & 50 & 995.6                         & 3.1289*$10^{-1}$ & 3.182*$10^{3}$  & 6.52*$10^{-9}$  & -1.88*$10^{4}$ & 1.051685 \\
		119 & tin-116          & 50 & 988.7                         & 3.1314*$10^{-1}$ & 3.157*$10^{3}$  & 6.75*$10^{-9}$  & -1.86*$10^{4}$ & 1.051775 \\
		120 & tin-115          & 50 & 979.1                         & 3.1343*$10^{-1}$ & 3.124*$10^{3}$  & 7.01*$10^{-9}$  & -1.84*$10^{4}$ & 1.051878 \\
		121 & cadmium-114      & 48 & 972.6                         & 3.1369*$10^{-1}$ & 3.1*$10^{3}$    & 7.25*$10^{-9}$  & -1.82*$10^{4}$ & 1.051969 \\
		122 & tin-114          & 50 & 971.6                         & 3.1371*$10^{-1}$ & 3.097*$10^{3}$  & 7.26*$10^{-9}$  & -1.82*$10^{4}$ & 1.051973 \\
		123 & indium-113       & 49 & 963.1                         & 3.1399*$10^{-1}$ & 3.067*$10^{3}$  & 7.54*$10^{-9}$  & -1.8*$10^{4}$  & 1.052074 \\
		124 & cadmium-112      & 48 & 957.0                         & 3.1425*$10^{-1}$ & 3.045*$10^{3}$  & 7.79*$10^{-9}$  & -1.79*$10^{4}$ & 1.052165 \\
		125 & tin-112          & 50 & 953.5                         & 3.1429*$10^{-1}$ & 3.034*$10^{3}$  & 7.84*$10^{-9}$  & -1.78*$10^{4}$ & 1.05218  \\
		126 & cadmium-111      & 48 & 947.6                         & 3.1455*$10^{-1}$ & 3.013*$10^{3}$  & 8.1*$10^{-9}$   & -1.77*$10^{4}$ & 1.052272 \\
		127 & cadmium-110      & 48 & 940.6                         & 3.1483*$10^{-1}$ & 2.988*$10^{3}$  & 8.4*$10^{-9}$   & -1.75*$10^{4}$ & 1.052369 \\
		128 & palladium-110    & 46 & 940.2                         & 3.1483*$10^{-1}$ & 2.986*$10^{3}$  & 8.4*$10^{-9}$   & -1.75*$10^{4}$ & 1.052371 \\
		129 & silver-109       & 47 & 931.7                         & 3.1513*$10^{-1}$ & 2.957*$10^{3}$  & 8.73*$10^{-9}$  & -1.73*$10^{4}$ & 1.052477 \\
		130 & palladium-108    & 46 & 925.2                         & 3.1541*$10^{-1}$ & 2.933*$10^{3}$  & 9.05*$10^{-9}$  & -1.71*$10^{4}$ & 1.052574 \\
		131 & cadmium-108      & 48 & 923.4                         & 3.1543*$10^{-1}$ & 2.927*$10^{3}$  & 9.08*$10^{-9}$  & -1.71*$10^{4}$ & 1.052583 \\
		132 & silver-107       & 47 & 915.3                         & 3.1574*$10^{-1}$ & 2.899*$10^{3}$  & 9.43*$10^{-9}$  & -1.69*$10^{4}$ & 1.05269  \\
		133 & palladium-106    & 46 & 909.5                         & 3.1601*$10^{-1}$ & 2.878*$10^{3}$  & 9.77*$10^{-9}$  & -1.68*$10^{4}$ & 1.052787 \\
		134 & cadmium-106      & 48 & 905.1                         & 3.1607*$10^{-1}$ & 2.864*$10^{3}$  & 9.84*$10^{-9}$  & -1.67*$10^{4}$ & 1.052808 \\
		135 & palladium-105    & 46 & 899.9                         & 3.1634*$10^{-1}$ & 2.845*$10^{3}$  & 1.02*$10^{-8}$  & -1.66*$10^{4}$ & 1.052903 \\
		136 & ruthenium-104    & 44 & 893.1                         & 3.1663*$10^{-1}$ & 2.821*$10^{3}$  & 1.06*$10^{-8}$  & -1.64*$10^{4}$ & 1.053008 \\
		137 & palladium-104    & 46 & 892.8                         & 3.1663*$10^{-1}$ & 2.82*$10^{3}$   & 1.06*$10^{-8}$  & -1.64*$10^{4}$ & 1.053009 \\
		138 & rhodium-103      & 45 & 884.2                         & 3.1696*$10^{-1}$ & 2.79*$10^{3}$   & 1.1*$10^{-8}$   & -1.62*$10^{4}$ & 1.053125 \\
		139 & ruthenium-102    & 44 & 877.9                         & 3.1725*$10^{-1}$ & 2.767*$10^{3}$  & 1.14*$10^{-8}$  & -1.61*$10^{4}$ & 1.053229 \\
		140 & palladium-102    & 46 & 875.2                         & 3.1729*$10^{-1}$ & 2.758*$10^{3}$  & 1.15*$10^{-8}$  & -1.6*$10^{4}$  & 1.053243 \\
		141 & ruthenium-101    & 44 & 868.7                         & 3.1759*$10^{-1}$ & 2.735*$10^{3}$  & 1.19*$10^{-8}$  & -1.58*$10^{4}$ & 1.05335  \\
		142 & ruthenium-100    & 44 & 861.9                         & 3.179*$10^{-1}$  & 2.711*$10^{3}$  & 1.24*$10^{-8}$  & -1.57*$10^{4}$ & 1.053461 \\
		143 & ruthenium-99     & 44 & 852.3                         & 3.1826*$10^{-1}$ & 2.678*$10^{3}$  & 1.3*$10^{-8}$   & -1.55*$10^{4}$ & 1.053588 \\
		144 & molybdenum-98    & 42 & 846.2                         & 3.1856*$10^{-1}$ & 2.656*$10^{3}$  & 1.35*$10^{-8}$  & -1.53*$10^{4}$ & 1.053698 \\
		145 & ruthenium-98     & 44 & 844.8                         & 3.1858*$10^{-1}$ & 2.652*$10^{3}$  & 1.35*$10^{-8}$  & -1.53*$10^{4}$ & 1.053705 \\
		146 & molybdenum-97    & 42 & 837.6                         & 3.1891*$10^{-1}$ & 2.626*$10^{3}$  & 1.41*$10^{-8}$  & -1.51*$10^{4}$ & 1.053823 \\
		147 & molybdenum-96    & 42 & 830.8                         & 3.1924*$10^{-1}$ & 2.602*$10^{3}$  & 1.46*$10^{-8}$  & -1.5*$10^{4}$  & 1.05394  \\
		148 & ruthenium-96     & 44 & 826.5                         & 3.193*$10^{-1}$  & 2.588*$10^{3}$  & 1.48*$10^{-8}$  & -1.49*$10^{4}$ & 1.053964 \\
		149 & molybdenum-95    & 42 & 821.6                         & 3.196*$10^{-1}$  & 2.571*$10^{3}$  & 1.53*$10^{-8}$  & -1.48*$10^{4}$ & 1.054072 \\
		150 & zirconium-94     & 40 & 814.7                         & 3.1994*$10^{-1}$ & 2.546*$10^{3}$  & 1.6*$10^{-8}$   & -1.46*$10^{4}$ & 1.054193 \\
		151 & molybdenum-94    & 42 & 814.3                         & 3.1995*$10^{-1}$ & 2.545*$10^{3}$  & 1.6*$10^{-8}$   & -1.46*$10^{4}$ & 1.054196 \\
		152 & niobium-93       & 41 & 805.8                         & 3.2031*$10^{-1}$ & 2.516*$10^{3}$  & 1.67*$10^{-8}$  & -1.44*$10^{4}$ & 1.054328 \\
		153 & zirconium-92     & 40 & 799.7                         & 3.2065*$10^{-1}$ & 2.494*$10^{3}$  & 1.74*$10^{-8}$  & -1.43*$10^{4}$ & 1.054448 \\
		154 & molybdenum-92    & 42 & 796.5                         & 3.207*$10^{-1}$  & 2.484*$10^{3}$  & 1.75*$10^{-8}$  & -1.42*$10^{4}$ & 1.054467 \\
		155 & zirconium-91     & 40 & 791.1                         & 3.2103*$10^{-1}$ & 2.464*$10^{3}$  & 1.82*$10^{-8}$  & -1.41*$10^{4}$ & 1.054585 \\
		156 & zirconium-90     & 40 & 783.9                         & 3.2139*$10^{-1}$ & 2.439*$10^{3}$  & 1.9*$10^{-8}$   & -1.39*$10^{4}$ & 1.054716 \\
		157 & yttrium-89       & 39 & 775.5                         & 3.2177*$10^{-1}$ & 2.41*$10^{3}$   & 2.*$10^{-8}$    & -1.38*$10^{4}$ & 1.054855 \\
		158 & strontium-88     & 38 & 768.5                         & 3.2214*$10^{-1}$ & 2.386*$10^{3}$  & 2.09*$10^{-8}$  & -1.36*$10^{4}$ & 1.054989 \\
		159 & strontium-87     & 38 & 757.4                         & 3.2258*$10^{-1}$ & 2.348*$10^{3}$  & 2.2*$10^{-8}$   & -1.34*$10^{4}$ & 1.055151 \\
		160 & krypton-86       & 36 & 749.2                         & 3.2298*$10^{-1}$ & 2.32*$10^{3}$   & 2.31*$10^{-8}$  & -1.32*$10^{4}$ & 1.055297 \\
		161 & strontium-86     & 38 & 748.9                         & 3.2299*$10^{-1}$ & 2.319*$10^{3}$  & 2.31*$10^{-8}$  & -1.32*$10^{4}$ & 1.055299 \\
		162 & rubidium-85      & 37 & 739.3                         & 3.2342*$10^{-1}$ & 2.286*$10^{3}$  & 2.43*$10^{-8}$  & -1.3*$10^{4}$  & 1.055457 \\
		163 & krypton-84       & 36 & 732.3                         & 3.2381*$10^{-1}$ & 2.261*$10^{3}$  & 2.55*$10^{-8}$  & -1.28*$10^{4}$ & 1.0556   \\
		164 & strontium-84     & 38 & 728.9                         & 3.2388*$10^{-1}$ & 2.251*$10^{3}$  & 2.57*$10^{-8}$  & -1.27*$10^{4}$ & 1.055623 \\
		165 & krypton-83       & 36 & 721.7                         & 3.2428*$10^{-1}$ & 2.226*$10^{3}$  & 2.69*$10^{-8}$  & -1.26*$10^{4}$ & 1.05577  \\
		166 & krypton-82       & 36 & 714.3                         & 3.2469*$10^{-1}$ & 2.2*$10^{3}$    & 2.82*$10^{-8}$  & -1.24*$10^{4}$ & 1.055921 \\
		167 & bromine-81       & 35 & 704.4                         & 3.2516*$10^{-1}$ & 2.166*$10^{3}$  & 2.98*$10^{-8}$  & -1.22*$10^{4}$ & 1.056093 \\
		168 & selenium-80      & 34 & 696.9                         & 3.2558*$10^{-1}$ & 2.14*$10^{3}$   & 3.14*$10^{-8}$  & -1.2*$10^{4}$  & 1.05625  \\
		169 & krypton-80       & 36 & 695.4                         & 3.2561*$10^{-1}$ & 2.136*$10^{3}$  & 3.15*$10^{-8}$  & -1.2*$10^{4}$  & 1.056261 \\
		170 & bromine-79       & 35 & 686.3                         & 3.2608*$10^{-1}$ & 2.105*$10^{3}$  & 3.32*$10^{-8}$  & -1.18*$10^{4}$ & 1.056433 \\
		171 & selenium-78      & 34 & 680.0                         & 3.265*$10^{-1}$  & 2.083*$10^{3}$  & 3.49*$10^{-8}$  & -1.17*$10^{4}$ & 1.056588 \\
		172 & krypton-78       & 36 & 675.6                         & 3.2659*$10^{-1}$ & 2.069*$10^{3}$  & 3.52*$10^{-8}$  & -1.16*$10^{4}$ & 1.056621 \\
		173 & selenium-77      & 34 & 669.5                         & 3.2701*$10^{-1}$ & 2.047*$10^{3}$  & 3.7*$10^{-8}$   & -1.15*$10^{4}$ & 1.056777 \\
		174 & selenium-76      & 34 & 662.1                         & 3.2747*$10^{-1}$ & 2.022*$10^{3}$  & 3.9*$10^{-8}$   & -1.13*$10^{4}$ & 1.056947 \\
		175 & arsenic-75       & 33 & 652.6                         & 3.2797*$10^{-1}$ & 1.99*$10^{3}$   & 4.13*$10^{-8}$  & -1.11*$10^{4}$ & 1.057136 \\
		176 & germanium-74     & 32 & 645.7                         & 3.2844*$10^{-1}$ & 1.966*$10^{3}$  & 4.35*$10^{-8}$  & -1.09*$10^{4}$ & 1.057308 \\
		177 & selenium-74      & 34 & 642.9                         & 3.285*$10^{-1}$  & 1.957*$10^{3}$  & 4.38*$10^{-8}$  & -1.09*$10^{4}$ & 1.05733  \\
		178 & germanium-73     & 32 & 635.5                         & 3.2898*$10^{-1}$ & 1.932*$10^{3}$  & 4.63*$10^{-8}$  & -1.07*$10^{4}$ & 1.05751  \\
		179 & germanium-72     & 32 & 628.7                         & 3.2945*$10^{-1}$ & 1.908*$10^{3}$  & 4.88*$10^{-8}$  & -1.06*$10^{4}$ & 1.057688 \\
		180 & gallium-71       & 31 & 619.0                         & 3.3001*$10^{-1}$ & 1.876*$10^{3}$  & 5.2*$10^{-8}$   & -1.04*$10^{4}$ & 1.057895 \\
		181 & zinc-70          & 30 & 611.1                         & 3.3053*$10^{-1}$ & 1.849*$10^{3}$  & 5.51*$10^{-8}$  & -1.02*$10^{4}$ & 1.05809  \\
		182 & germanium-70     & 32 & 610.5                         & 3.3054*$10^{-1}$ & 1.847*$10^{3}$  & 5.52*$10^{-8}$  & -1.02*$10^{4}$ & 1.058095 \\
		183 & gallium-69       & 31 & 602.0                         & 3.3108*$10^{-1}$ & 1.818*$10^{3}$  & 5.86*$10^{-8}$  & -1.*$10^{4}$   & 1.0583   \\
		184 & zinc-68          & 30 & 595.4                         & 3.3159*$10^{-1}$ & 1.796*$10^{3}$  & 6.2*$10^{-8}$   & -9.88*$10^{3}$ & 1.058492 \\
		185 & zinc-67          & 30 & 585.2                         & 3.322*$10^{-1}$  & 1.762*$10^{3}$  & 6.63*$10^{-8}$  & -9.67*$10^{3}$ & 1.058722 \\
		186 & zinc-66          & 30 & 578.1                         & 3.3274*$10^{-1}$ & 1.737*$10^{3}$  & 7.03*$10^{-8}$  & -9.52*$10^{3}$ & 1.058927 \\
		187 & copper-65        & 29 & 569.2                         & 3.3334*$10^{-1}$ & 1.708*$10^{3}$  & 7.51*$10^{-8}$  & -9.34*$10^{3}$ & 1.059155 \\
		188 & nickel-64        & 28 & 561.8                         & 3.3392*$10^{-1}$ & 1.682*$10^{3}$  & 7.99*$10^{-8}$  & -9.18*$10^{3}$ & 1.059374 \\
		189 & zinc-64          & 30 & 559.1                         & 3.3399*$10^{-1}$ & 1.674*$10^{3}$  & 8.05*$10^{-8}$  & -9.13*$10^{3}$ & 1.0594   \\
		190 & copper-63        & 29 & 551.4                         & 3.3458*$10^{-1}$ & 1.648*$10^{3}$  & 8.58*$10^{-8}$  & -8.97*$10^{3}$ & 1.059627 \\
		191 & nickel-62        & 28 & 545.3                         & 3.3514*$10^{-1}$ & 1.627*$10^{3}$  & 9.12*$10^{-8}$  & -8.84*$10^{3}$ & 1.059842 \\
		192 & nickel-61        & 28 & 534.7                         & 3.3584*$10^{-1}$ & 1.592*$10^{3}$  & 9.82*$10^{-8}$  & -8.63*$10^{3}$ & 1.060111 \\
		193 & nickel-60        & 28 & 526.8                         & 3.3648*$10^{-1}$ & 1.566*$10^{3}$  & 1.05*$10^{-7}$  & -8.47*$10^{3}$ & 1.060355 \\
		194 & cobalt-59        & 27 & 517.3                         & 3.3718*$10^{-1}$ & 1.534*$10^{3}$  & 1.13*$10^{-7}$  & -8.27*$10^{3}$ & 1.060626 \\
		195 & iron-58          & 26 & 509.9                         & 3.3784*$10^{-1}$ & 1.509*$10^{3}$  & 1.21*$10^{-7}$  & -8.12*$10^{3}$ & 1.060878 \\
		196 & nickel-58        & 28 & 506.5                         & 3.3794*$10^{-1}$ & 1.499*$10^{3}$  & 1.22*$10^{-7}$  & -8.06*$10^{3}$ & 1.060919 \\
		197 & iron-57          & 26 & 499.9                         & 3.3859*$10^{-1}$ & 1.476*$10^{3}$  & 1.31*$10^{-7}$  & -7.92*$10^{3}$ & 1.061169 \\
		198 & iron-56          & 26 & 492.3                         & 3.3928*$10^{-1}$ & 1.451*$10^{3}$  & 1.41*$10^{-7}$  & -7.77*$10^{3}$ & 1.061439 \\
		199 & manganese-55     & 25 & 482.1                         & 3.4007*$10^{-1}$ & 1.418*$10^{3}$  & 1.53*$10^{-7}$  & -7.57*$10^{3}$ & 1.061748 \\
		200 & chromium-54      & 24 & 474.0                         & 3.4082*$10^{-1}$ & 1.391*$10^{3}$  & 1.65*$10^{-7}$  & -7.4*$10^{3}$  & 1.062038 \\
		201 & iron-54          & 26 & 471.8                         & 3.4089*$10^{-1}$ & 1.384*$10^{3}$  & 1.66*$10^{-7}$  & -7.36*$10^{3}$ & 1.062067 \\
		202 & chromium-53      & 24 & 464.3                         & 3.4163*$10^{-1}$ & 1.359*$10^{3}$  & 1.79*$10^{-7}$  & -7.21*$10^{3}$ & 1.062359 \\
		203 & chromium-52      & 24 & 456.3                         & 3.4241*$10^{-1}$ & 1.333*$10^{3}$  & 1.94*$10^{-7}$  & -7.05*$10^{3}$ & 1.062665 \\
		204 & vanadium-51      & 23 & 445.8                         & 3.433*$10^{-1}$  & 1.299*$10^{3}$  & 2.12*$10^{-7}$  & -6.85*$10^{3}$ & 1.063017 \\
		205 & titanium-50      & 22 & 437.8                         & 3.4413*$10^{-1}$ & 1.272*$10^{3}$  & 2.3*$10^{-7}$   & -6.69*$10^{3}$ & 1.063345 \\
		206 & chromium-50      & 24 & 435.0                         & 3.4423*$10^{-1}$ & 1.264*$10^{3}$  & 2.32*$10^{-7}$  & -6.65*$10^{3}$ & 1.063385 \\
		207 & titanium-49      & 22 & 426.8                         & 3.4509*$10^{-1}$ & 1.237*$10^{3}$  & 2.52*$10^{-7}$  & -6.48*$10^{3}$ & 1.063726 \\
		208 & titanium-48      & 22 & 418.7                         & 3.4597*$10^{-1}$ & 1.21*$10^{3}$   & 2.75*$10^{-7}$  & -6.32*$10^{3}$ & 1.064077 \\
		209 & titanium-47      & 22 & 407.1                         & 3.4702*$10^{-1}$ & 1.173*$10^{3}$  & 3.04*$10^{-7}$  & -6.11*$10^{3}$ & 1.064496 \\
		210 & calcium-46       & 20 & 398.8                         & 3.4796*$10^{-1}$ & 1.146*$10^{3}$  & 3.33*$10^{-7}$  & -5.95*$10^{3}$ & 1.064875 \\
		211 & titanium-46      & 22 & 398.2                         & 3.4799*$10^{-1}$ & 1.144*$10^{3}$  & 3.34*$10^{-7}$  & -5.94*$10^{3}$ & 1.064885 \\
		212 & scandium-45      & 21 & 387.8                         & 3.4905*$10^{-1}$ & 1.111*$10^{3}$  & 3.7*$10^{-7}$   & -5.74*$10^{3}$ & 1.065313 \\
		213 & calcium-44       & 20 & 381.0                         & 3.4999*$10^{-1}$ & 1.088*$10^{3}$  & 4.04*$10^{-7}$  & -5.61*$10^{3}$ & 1.065694 \\
		214 & calcium-43       & 20 & 369.8                         & 3.5116*$10^{-1}$ & 1.053*$10^{3}$  & 4.51*$10^{-7}$  & -5.4*$10^{3}$  & 1.066168 \\
		215 & calcium-42       & 20 & 361.9                         & 3.5221*$10^{-1}$ & 1.027*$10^{3}$  & 4.97*$10^{-7}$  & -5.25*$10^{3}$ & 1.0666   \\
		216 & potassium-41     & 19 & 351.6                         & 3.5342*$10^{-1}$ & 9.949*$10^{2}$  & 5.55*$10^{-7}$  & -5.06*$10^{3}$ & 1.067095 \\
		217 & argon-40         & 18 & 343.8                         & 3.5455*$10^{-1}$ & 9.697*$10^{2}$  & 6.15*$10^{-7}$  & -4.92*$10^{3}$ & 1.067559 \\
		218 & calcium-40       & 20 & 342.1                         & 3.5464*$10^{-1}$ & 9.645*$10^{2}$  & 6.2*$10^{-7}$   & -4.89*$10^{3}$ & 1.067596 \\
		219 & potassium-39     & 19 & 333.7                         & 3.5583*$10^{-1}$ & 9.379*$10^{2}$  & 6.9*$10^{-7}$   & -4.73*$10^{3}$ & 1.068092 \\
		220 & argon-38         & 18 & 327.3                         & 3.5696*$10^{-1}$ & 9.17*$10^{2}$   & 7.64*$10^{-7}$  & -4.61*$10^{3}$ & 1.068562 \\
		221 & chlorine-37      & 17 & 317.1                         & 3.5836*$10^{-1}$ & 8.849*$10^{2}$  & 8.63*$10^{-7}$  & -4.43*$10^{3}$ & 1.069144 \\
		222 & sulfur-36        & 16 & 308.7                         & 3.597*$10^{-1}$  & 8.583*$10^{2}$  & 9.7*$10^{-7}$   & -4.27*$10^{3}$ & 1.069707 \\
		223 & argon-36         & 18 & 306.7                         & 3.5982*$10^{-1}$ & 8.524*$10^{2}$  & 9.81*$10^{-7}$  & -4.24*$10^{3}$ & 1.069758 \\
		224 & chlorine-35      & 17 & 298.2                         & 3.6123*$10^{-1}$ & 8.256*$10^{2}$  & 1.11*$10^{-6}$  & -4.09*$10^{3}$ & 1.070352 \\
		225 & sulfur-34        & 16 & 291.8                         & 3.6255*$10^{-1}$ & 8.05*$10^{2}$   & 1.24*$10^{-6}$  & -3.97*$10^{3}$ & 1.070915 \\
		226 & sulfur-33        & 16 & 280.4                         & 3.6427*$10^{-1}$ & 7.698*$10^{2}$  & 1.43*$10^{-6}$  & -3.77*$10^{3}$ & 1.071651 \\
		227 & sulfur-32        & 16 & 271.8                         & 3.6588*$10^{-1}$ & 7.428*$10^{2}$  & 1.64*$10^{-6}$  & -3.62*$10^{3}$ & 1.072343 \\
		228 & phosphorus-31    & 15 & 262.9                         & 3.6758*$10^{-1}$ & 7.153*$10^{2}$  & 1.88*$10^{-6}$  & -3.47*$10^{3}$ & 1.073079 \\
		229 & silicon-30       & 14 & 255.6                         & 3.6924*$10^{-1}$ & 6.923*$10^{2}$  & 2.15*$10^{-6}$  & -3.34*$10^{3}$ & 1.073802 \\
		230 & silicon-29       & 14 & 245.0                         & 3.7126*$10^{-1}$ & 6.599*$10^{2}$  & 2.52*$10^{-6}$  & -3.16*$10^{3}$ & 1.074685 \\
		231 & silicon-28       & 14 & 236.5                         & 3.732*$10^{-1}$  & 6.338*$10^{2}$  & 2.94*$10^{-6}$  & -3.01*$10^{3}$ & 1.075542 \\
		232 & aluminum-27      & 13 & 225.0                         & 3.7553*$10^{-1}$ & 5.99*$10^{2}$   & 3.52*$10^{-6}$  & -2.82*$10^{3}$ & 1.076583 \\
		233 & magnesium-26     & 12 & 216.7                         & 3.7769*$10^{-1}$ & 5.737*$10^{2}$  & 4.14*$10^{-6}$  & -2.69*$10^{3}$ & 1.077551 \\
		234 & magnesium-25     & 12 & 205.6                         & 3.8027*$10^{-1}$ & 5.406*$10^{2}$  & 5.02*$10^{-6}$  & -2.51*$10^{3}$ & 1.078722 \\
		235 & magnesium-24     & 12 & 198.3                         & 3.826*$10^{-1}$  & 5.182*$10^{2}$  & 5.95*$10^{-6}$  & -2.39*$10^{3}$ & 1.07979  \\
		236 & sodium-23        & 11 & 186.6                         & 3.8561*$10^{-1}$ & 4.838*$10^{2}$  & 7.37*$10^{-6}$  & -2.2*$10^{3}$  & 1.081181 \\
		237 & neon-22          & 10 & 177.8                         & 3.8847*$10^{-1}$ & 4.576*$10^{2}$  & 9.*$10^{-6}$    & -2.07*$10^{3}$ & 1.08252  \\
		238 & neon-21          & 10 & 167.4                         & 3.9178*$10^{-1}$ & 4.273*$10^{2}$  & 1.13*$10^{-5}$  & -1.91*$10^{3}$ & 1.084081 \\
		239 & neon-20          & 10 & 160.6                         & 3.9479*$10^{-1}$ & 4.069*$10^{2}$  & 1.38*$10^{-5}$  & -1.8*$10^{3}$  & 1.085525 \\
		240 & fluorine-19      & 9  & 147.8                         & 3.9907*$10^{-1}$ & 3.704*$10^{2}$  & 1.82*$10^{-5}$  & -1.61*$10^{3}$ & 1.0876   \\
		241 & oxygen-18        & 8  & 139.8                         & 4.0288*$10^{-1}$ & 3.47*$10^{2}$   & 2.31*$10^{-5}$  & -1.49*$10^{3}$ & 1.089478 \\
		242 & oxygen-17        & 8  & 131.8                         & 4.0704*$10^{-1}$ & 3.237*$10^{2}$  & 2.98*$10^{-5}$  & -1.37*$10^{3}$ & 1.091564 \\
		243 & oxygen-16        & 8  & 127.6                         & 4.1074*$10^{-1}$ & 3.107*$10^{2}$  & 3.72*$10^{-5}$  & -1.3*$10^{3}$  & 1.093442 \\
		244 & nitrogen-15      & 7  & 115.5                         & 4.1673*$10^{-1}$ & 2.771*$10^{2}$  & 5.24*$10^{-5}$  & -1.14*$10^{3}$ & 1.096538 \\
		245 & nitrogen-14      & 7  & 104.7                         & 4.2319*$10^{-1}$ & 2.473*$10^{2}$  & 7.48*$10^{-5}$  & -9.94*$10^{2}$ & 1.099958 \\
		246 & carbon-13        & 6  & 97.11                         & 4.2948*$10^{-1}$ & 2.261*$10^{2}$  & 1.04*$10^{-4}$  & -8.9*$10^{2}$  & 1.103371 \\
		247 & carbon-12        & 6  & 92.16                         & 4.3565*$10^{-1}$ & 2.116*$10^{2}$  & 1.42*$10^{-4}$  & -8.16*$10^{2}$ & 1.106795 \\
		248 & boron-11         & 5  & 76.20                         & 4.4753*$10^{-1}$ & 1.703*$10^{2}$  & 2.5*$10^{-4}$   & -6.32*$10^{2}$ & 1.113615 \\
		249 & boron-10         & 5  & 64.75                         & 4.5996*$10^{-1}$ & 1.408*$10^{2}$  & 4.32*$10^{-4}$  & -5.02*$10^{2}$ & 1.121078 \\
		250 & beryllium-9      & 4  & 58.16                         & 4.7184*$10^{-1}$ & 1.233*$10^{2}$  & 7.01*$10^{-4}$  & -4.22*$10^{2}$ & 1.128538 \\
		251 & lithium-7        & 3  & 39.24                         & 5.1231*$10^{-1}$ & 7.66*$10^{1}$   & 2.9*$10^{-3}$   & -2.29*$10^{2}$ & 1.156422 \\
		252 & lithium-6        & 3  & 31.99                         & 5.4087*$10^{-1}$ & 5.915*$10^{1}$  & 6.64*$10^{-3}$  & -1.6*$10^{2}$  & 1.178496 \\
		253 & helium-4         & 2  & 28.30                         & 6.0521*$10^{-1}$ & 4.675*$10^{1}$  & 2.98*$10^{-2}$  & -9.94*$10^{1}$ & 1.235472 \\
		254 & deuterium        & 1  & 2.225                         & 2.6432                          & 8.416*$10^{-1}$ & 1.42*$10^{5}$   & 2.64*$10^{1}$  & 5.069582
\end{longtable}
\begin{figure}[h!]
	\includegraphics[draft=false]{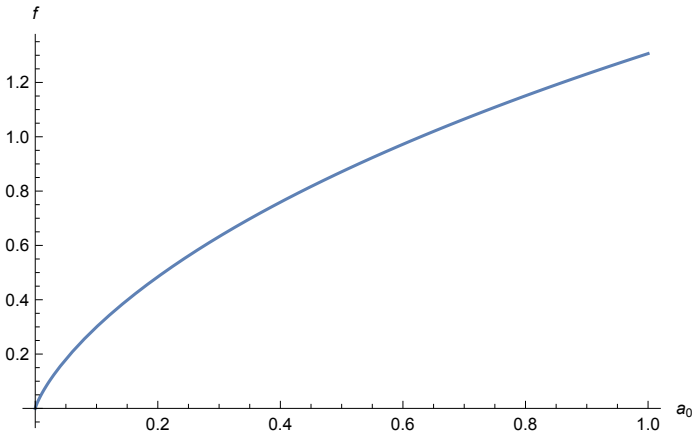}
	\caption{The dependence of the left hand side of the  equation ~\eqref{NTm}  on $a_0$}
	\label{masl_fig_4}
\end{figure}

Here  $\mathcal{Z}$ is the charge number (number of protons) in the nucleus and  $E_b$ is the binding energy of the nucleus, equal to the core temperature $T$. It is well known that  the binding energy of a nucleus increases with increasing mass number $A$. At the same time,  the mass of the nucleus $m$ and its volume $V$ also  grow.
The equation \eqref{N=0} for $\gamma=1/2$ can be written as follows:
\begin{equation}
	\label{NTm}
	\frac{1}{2} \operatorname{Li}_{3/2}(a_0)-  \log (a_0)\operatorname{Li}_{1/2}(a_0)=\frac{1}{V}\left(\frac{2\pi\hbar^2}{m T}\right)^{3/2}.
\end{equation}

The value   $a=a_0$  is the minimum value of activity for the Bose branch, since for  $a<a_0$ the region of uncertainty   occurs  where $N\equiv0$  and $P\equiv0$  for all  $a$  up to the  value of  $a=0$. Here the passage  to the Fermi  branch  begins. Hence  $a_0$  is  the point  separating the  Bose  case  from  the  case of  uncertainty.

The left hand side   of Eq.~\eqref{NTm}  which we denote  by $f(a_0)$  decreases  when   the value of  $a_0$  decreases (see Fig.~\ref{masl_fig_4}).
The  value in the right hand side  of~\eqref{NTm} also decreases with the growth of the nucleus mass number~$A$.

Therefore,  as one  can see from   the  Table,  the activity $a_0$ decreases  and the absolute value of the chemical potential $|\mu_0|=T|\log{a_0}|$ increases when the mass number $A$   increases.

 The variance of the particle number~$\overline{(\Delta N)^{2}}$  for  given variables  $\mu,V, T$ in  a grand canonical ensemble is determined  by the relation (see~\cite{Kvasnikov-3}):
\begin{equation}\label{disN}
\overline{(\Delta N)^{2}}=T\left(\frac{\partial {N}}{\partial \mu}\right)_{T}.
\end{equation}

By differentiating with respect to $\mu$  the both sides of the relation 
\begin{equation}
N(\mu,V,T)= \frac{V}{\lambda^{D}} (\operatorname{Li}_{1+\gamma}(a)-\frac{1}{(N(\mu,V,T)+1)^{\gamma}}\operatorname{Li}_{1+\gamma}(a^{N(\mu,V,T)+1})),
\end{equation}
we obtain   the  relation for $\left(\frac{\partial {N}}{\partial \mu}\right)_{T}$:
\begin{equation}\label{dnmu}
\left(\frac{\partial {N}}{\partial \mu}\right)_{T}=\frac{(N+1)^{\gamma+1} \left(-(N+1)^{1-\gamma} \text{Li}_{\gamma }\left(\left(e^{\mu /T}\right)^{N+1}\right)+\text{Li}_{\gamma }\left(e^{\mu /T}\right)\right)}{T \left(\frac{\lambda^{\gamma+1}}{V}(N+1)^{\gamma +1}-\gamma  \text{Li}_{\gamma +1}\left(\left(e^{\mu /T}\right)^{N+1}\right)+(N+1) \log \left(e^{\mu /T}\right) \text{Li}_{\gamma }\left(\left(e^{\mu /T}\right)^{N+1}\right)\right)},
\end{equation}
where  $N=N(\mu,V,T)$.

The expression for  dispersion for small   $N$  in the zeroth approximation has  the  form
\begin{equation}\label{dnmuas}
\overline{(\Delta N)^{2}}=-{2}\frac{ (1-\gamma)  \text{Li}_{\gamma }(a_0)+\log (a_0) \text{Li}_{\gamma -1}(a_0)}{\log ^2(a) \text{Li}_{\gamma-1}(a_0)+\gamma ((\gamma+1) \text{Li}_{\gamma+1}(a_0)-2 \log (a_0) \text{Li}_\gamma(a_0))}.
\end{equation}

Let us denote for arbitrary value  $x$    its  mean square fluctuation  by  $\delta x=\sqrt{\overline{(\Delta x)^{2}}}.$
Using  a  well-known for a grand canonical ensemble  relation of uncertainty  $\delta N \delta \mu\ge T$,
one can find  the minimum value of  mean square fluctuation  $ \delta \mu_{min}=T/{\delta N}\leq \delta \mu$.

The values  $\delta \mu_{min}$ (in MeV)  and  $\delta N$ are presented on the Table  above.

The  value $\delta \mu$   determines the halo width and  monotonically changes  with all columns of  the  Table.

Thus  we have obtained  a new  table  for nuclear  physics  which   demonstrates a monotonic relation between   the nucleus mass number~$A$, the binding energy  $E_b$, the minimum value of activity $a=a_0$,  the chemical  potential $\mu_0=T\log{a_0}$,  the compressibility  factor $F={PV}/{NT}$ for $a_0$,  and  also    the minimum value of  mean square fluctuation  $ \delta \mu_{min}=T/{\delta N}\leq \delta \mu$.  The values $\delta \mu_{min}$  and  $\delta N$  are involved  in   uncertainty relations of nuclear physics.

\end{document}